\documentclass[
aps,
prd,
superscriptaddress,secnumarabic,
amssymb,amsmath,nobibnotes,aps,prd,showkeys,showpacs,nofootinbib
]{revtex4-2}
\usepackage{graphicx}
\usepackage{sidecap}
\usepackage[hang,small,it]{caption}
\usepackage{amsmath}
\usepackage{multido}
\usepackage{romannum}
\makeatletter
\ams@newcommand{\vardot}[2]{%
	{\mathop{#2\kern0pt}\limits^{\vbox to-1.4\ex@{\kern-\tw@\ex@
				\hbox{\normalfont\multido{}{#1}{.}}\vss}}}}
\makeatother

\usepackage{hyperref}
\usepackage{amsmath,amsfonts,amssymb}
\hypersetup{dvips,dvipdfm,colorlinks=true,urlcolor=green,filecolor=magenta,linktoc=page,citecolor=red,linkcolor=blue,bookmarks=true}
\usepackage{graphicx,epstopdf}
\usepackage{dcolumn}   
\usepackage{subcaption}
\usepackage{multirow}
\usepackage{multirow,booktabs}
\usepackage{makecell}
\usepackage[english]{babel}
\usepackage{appendix}
\newtheorem{prop}{Proposition}

\begin{document}
	\title{A dynamical system analysis of bouncing cosmology with spatial curvature}

	\author{Soumya Chakraborty}
	\email{soumyachakraborty150@gmail.com}
	\affiliation{Department of Mathematics, Jadavpur University, Kolkata-700 032, West Bengal, India}
	\author{Sudip Mishra}
	\email[Corresponding author: ]{sudipcmiiitmath@gmail.com}
	\affiliation{Department of Applied Mathematics, Maulana Abul Kalam Azad University
		of Technology, Kolkata-700064 and Haringhata, Nadia-741249 (main campus), West Bengal, India}
	\author{Subenoy Chakraborty}
	\email{schakraborty.math@gmail.com}
	\affiliation{Department of Mathematics, Jadavpur University, Kolkata-700 032, West Bengal, India}

	\begin{abstract}
		
		The present work deals with a FLRW cosmological model with spatial curvature and minimally coupled scalar field as the matter content.  The curvature term behaves as a perfect fluid with the equation of state parameter $\omega_{\mathcal{K}}=-\frac{1}{3}$.  Using suitable transformation of variables, the evolution equations are reduced to an autonomous system for both power law and exponential form of the scalar potential.  The critical points are analyzed with center manifold theory and stability has been discussed.  Also, critical points at infinity have been studied using the notion of Poincar\'{e} sphere.  Finally, the cosmological implications of the critical points and cosmological bouncing scenarios are discussed.  It is found that the cosmological bounce takes place near the points at infinity when the non-isolated critical points on the equator of the Poincar\'{e} sphere are saddle or saddle-node in nature.
		
	\end{abstract}

	
	
	\maketitle	
	\newpage	
	\section{Introduction}

	The $\Lambda$CDM model is the simplest and very well-known cosmological model that describes the evolution of the universe from the early inflationary paradigm to the present era of expansion.  Despite being successful, according to the largest part of the cosmology community, in describing the formation and evolution of the large scale structure in the universe,
	$\Lambda$CDM model has a large family of challenges that have not been resolved. Also,  it is not consistent with many observations e.g. H0-tension \cite{PhysRevD.104.023523}, missing satellites \cite{galaxies5010017}, cosmological constant problems \cite{Cosmological_constraints}, the initial condition of the universe before the inflationary era  \cite{Dabrowski:1995ae, Graham:2011nb, Graham:2014pca, Graham:2014pca, Planck:2018vyg} etc.
	
	As the present observations are consistent with the expanding universe so it is reasonable to choose a model of the universe which keeps expanding from the very beginning till the present era \cite{Starobinsky:1999yw}.  However, from theoretical point of view, the universe may have an era of contraction in the very early epoch before inflation or the universe may experience a contracting phase at far future \cite{1978SvAL, PhysRevD.79.043524, LeonTorres:2010cdf}.  Friedmann equations tell us that for negative spatial curvature, the universe is an ever-expanding model, for flat model the expansion continues until the total energy of the universe is positive,  for positive spatial curvature the expansion will halt at a finite time and there will be a phase of contraction \cite{MATSUI2019152}.  Further, for a scalar field cosmology, there will have a contraction followed by an expanding era for a range of values of the scalar field.  Usually, in the contracting phase, the kinetic energy of the scalar field increases, and it stands to dominate the evolution.  Then there will be no mechanism to halt this collapse leading to a singularity \cite{Alam:2004jy}.  Though theoretically, Einstein field equations for spatial curvature indicates a possibility for contracting phase in the past or future, however, there is no observational hint for a possible contracting era.
	
	However, it will be interesting if by some mechanism the contraction can be halted and again there is expansion - the bouncing model universe.  If the bouncing phase occurs before the plank energy scale, then Einstein's gravity may be chosen as an effective field theory.  Further, the existence of the bouncing solution depends on the spatial curvature of the space-time.  For flat or open universe, bouncing phase is characterized by the null energy condition while for closed model of the universe, the bouncing era occurs when the curvature term balances the total energy of the universe.  Thus there are two essential ingredients for the bouncing scenario namely the positive spatial curvature and the scalar potential that gradually becomes flatter as the scalar field goes away from the minimum. 
	It is to be noted that the transition from expansion to contraction is possible due to positive spatial curvature.  Due to the suppression of the kinetic energy as the potential becomes sufficiently flat, the scalar field slow-rolls over a flat part of the potential and as a result, there is an accelerated contraction of the universe.  Thus for the occurrence of the bouncing scenario, the positive curvature term should balance the scalar field energy \cite{Battefeld:2014uga, Brandenberger:2016vhg, Ijjas:2018qbo, Singh:2018xjv, Agrawal:2021msm}.  However, from observational viewpoint, there is no indication of a bouncing phase of evolution- though $H$ is decreasing but it is positive from observational data.  The greatest advantage of bouncing scenario is avoidance of possible singularity either in past (Big Bang) or in future (Big crunch).
	
	The present work deals with scalar field cosmology in the background of FLRW space-time with positive spatial curvature.  The potential of the scalar field is chosen to be flatter as the scalar field goes away from the minimum.  One can induce a transition from expansion to contraction either by the spatial curvature.  The scalar field continues to climb up the potential after passing through the minimum potential and by converting most of the kinetic energy to the potential energy.  More specifically, for sufficiently flat potential (by suppressing kinetic energy) the scalar field experiences a slow roll on a flat plateau of the potential, and as a result, there will be an accelerated contracting phase of the universe.  Further, if the positive spatial curvature term is equivalent to the scalar field energy during this epoch then a bouncing scenario will occur.  Thus the accelerated contracting phase due to the flat potential is very much relevant for the bouncing universe \cite{MATSUI2019152, Bari_2019}.
	
	From an observational point of view the bouncing model is not a universally accepted model because positive curvature can at most be weakly supported by the latest CMB observation \cite{Graham:2011nb, Planck:2018vyg} but in addition, when lensing and baryon acoustic oscillation constraints are incorporated then the flat model is the best choice.  However, if future observations favor positive curvature then one may interpret it as the remnant of a contracting phase at the very early universe, firstly, it is to be mentioned that there are few works in the literature \cite{Adam:2011, Jow:2020} related to the cyclic universe where  positive spatial curvature were considered.  Instead of the scalar field, they have used domain walls for the requirement of the bouncing era. 
	
	The entirety of our work is swotted as follows: section \ref{BES} deals with the basic equations of the present cosmological model.  Also, the corresponding Einstein field equations, matter conservation equations, and the condition for the bouncing scenarios are also presented in this section.  With a suitable choice of the variables the evolution, equations are converted to an autonomous system in section \ref{FACS}.  The non-hyperbolic critical points are analyzed using center manifold theory and stability analysis has been discussed in the same section \ref{FACS}.   We expound compactification and dynamics around the points at infinity in section \ref{Infinity}.  We present cosmological implications in a section \ref{CI}.  Finally, the brief discussion and concluding remarks of the present work are proposed in section \ref{conclusion}.

	\section{Basic Equations\label{BES}}
	
	In the background of  Friedmann-Lema\^{i}tre-Robertson-Walker (FLRW) space-time with line element 
	\begin{align}
		\,ds^2=-\,dt^2+a^2(t)\left[\frac{\,dr^2}{1-\mathcal{K} r^2}+r^2(\,d\theta^2+\sin^2 \,d\Phi^2)\right],
	\end{align}	
	in co-moving coordinates $(t,r,\theta, \phi)$, 
	where $a(t)$ is the scale factor and $\mathcal{K}$ is the spatial curvature with $\mathcal{K}=0,1,-1$ corresponding to a flat, closed, or open universe, respectively.  We consider the Einstein-Hilbert action with a real scalar field $\phi$ \cite{MATSUI2019152},
	\begin{align}
		S=\int d^4x\sqrt{-g}\left(\frac{1}{2}R-\frac{1}{2}g^{\mu\nu}\partial_\mu\phi\partial_\nu\phi-V(\phi)\right),
	\end{align}	
	where $R$ is the Ricci scalar and $V(\phi)$ is the scalar potential.  Now the Friedmann equations are given by
	\begin{align}
		3\left(H^2+\frac{\mathcal{K}}{a^2}\right)&=\rho_\phi,\label{eq3}\\
		2\left(\dot{H}-\frac{\mathcal{K}}{a^2}\right)&=-(\rho_\phi+p_\phi),\label{eq4}
	\end{align}	
	where the `dot' denotes the derivative with respect to time, and $H=\frac{\dot{a}}{a}$ is the Hubble parameter.  Here $\rho_\phi$ and $p_\phi$ are energy density and pressure of the scalar field having expression
	\begin{align}
		\rho_\phi&=\frac{1}{2}\dot{\phi}^2+V(\phi),\label{eq5}\\
		p_\phi&=\frac{1}{2}\dot{\phi}^2-V(\phi).\label{eq6}
	\end{align}

	Now	the evolution of the scalar field $\phi$  can be written as \cite{MATSUI2019152}
		\begin{align}
			\ddot{\phi}+3H\dot{\phi}+V'(\phi)=0,\label{eq9}
		\end{align}	
		where $V'(\phi)$ denotes differentiation of the scalar potential $V(\phi)$ with respect to $\phi$.  Now the above field equations $(\ref{eq3}-\ref{eq4})$ can be rewritten as
		
		\begin{align}
			3H^2&=\rho_\phi+\rho_\mathcal{K},\label{eq7} \\ 
			2\dot{H}&=-(\rho_\phi+p_\phi)-(\rho_\mathcal{K}+p_{\mathcal{K})} \label{eq8}
		\end{align}	
	where
			\begin{align*}	
			\rho_\mathcal{K}=-\frac{3\mathcal{K}}{a^2},\quad \text{and}\quad
			p_\mathcal{K}=\frac{\mathcal{\mathcal{K}}}{a^2}
		\end{align*}	
		are energy density and pressure of the hypothetical curvature fluid having equation of state parameter $\omega_{\mathcal{K}}=-\frac{1}{3}$.  Thus the density parameter associated with the curvature fluid in the above first Friedmann equations is given by \cite{MATSUI2019152}
	$$
	\Omega_{\mathcal{K}}=\frac{\rho_{curv}}{\rho_{critical}}=-\frac{\mathcal{K}}{a^2H^2}.
	$$	
One may note from the first Friedmann equation (i.e., Eq.(\ref{eq3})) that for a given scalar field energy density if the positive curvature term is significant then it gradually tries to reduce the expansion of the universe.  The expansion can even be stopped if the curvature term balances the scalar field energy density.  However, from the second Friedmann equation (i.e., Eq.(\ref{eq4})), it is easy to see that at the very early era when `$a$' is very small then positive curvature term dominates over the kinetic energy of the scalar field and consequently $\dot{H}$ is positive (i.e., $H$ is increasing) and the universe is going through an expanding era.  But with the expansion, the scale factor gradually increases and at one stage the curvature term balances the kinetic energy term of the scalar field and $\dot{H}$ vanishes at that instant.  Subsequently, due to dominance of the kinetic energy of the scalar field $\dot{H}$ becomes negative.  So $H$ gradually decreases and then there are two possibilities (i) $H$ reaches a minimum (where $H>0$) and then again $H$ starts increasing indicating a bouncing era at minimum $H$, or (ii) $H$ gradually decreases, passing through zero and then becomes negative, indicating a contracting phase \cite{MATSUI2019152}.  This may be considered as an example how Einstein gravity may also describe a contracting phase for which so far there are no observational predictions.
	
	On the other hand, during the contracting phase, the scalar field experiences a negative friction contribution (anti-friction) and if $|H|$ is assumed to be small compared to the mass or characteristic curvature of the scalar potential.  Then the friction (i.e., second term in equation (\ref{eq9})) will be negligible compare to the acceleration term and the force term.  So, the evolution equation of the scalar field approximately becomes
	\begin{align}
	\ddot{\phi}+\frac{\partial V}{\partial \phi}\approx 0.\label{eqt10}
	\end{align}
	
In particular, if $V=\frac{1}{2}m^2\phi^2$ then from Eq.(\ref{eqt10}), $\ddot{\phi}+m^2\phi\approx 0$, that is, the scalar field oscillates like a simple harmonic oscillator.  Subsequently, when $|H|$ becomes comparable to (or larger than) the characteristic mass scale, then the kinetic term gradually grows rapidly and the universe enters the big crunch.  However, to avoid this ever-contracting evolution, one may note that during the simple harmonic motion of the scalar field, the kinetic term temporarily vanishes at the two endpoints.  If the potential has a flat nature at the endpoints with $V(\phi)>0$, then the evolution will be dominated by the potential energy for a longer time.  But as the kinetic term grows much faster so at some epoch it balances the potential term and also the positive curvature term and the Hubble parameter vanishes again.  This phase is known as the bouncing era.  After this bounce, the universe again starts expanding.  The well-known bouncing solution in cosmology is the de-Sitter bounce solution given by 
	$$
	a(t)=\sqrt{\frac{3\mathcal{K}}{\Lambda}}\cosh \left(\sqrt{\frac{\Lambda}{3}}t\right),
	$$
	where $\mathcal{K}>0$ is the spatial curvature and $\Lambda$ is the positive cosmological constant \cite{MATSUI2019152}.

	\section{Formation of the autonomous system: critical point and stability analysis\label{FACS}}
	We try to obtain a qualitative picture of this
	cosmological model, that's why we adopt a dynamical system approach and for that choose the following dynamical variables \cite{Bahamonde:2017ize, Dutta:2010yw, SavasArapoglu:2018tjk}
	\begin{align}
		x:&=\frac{\dot{\phi}}{\sqrt{6}H},\label{eqn11}\\
		y:&=\frac{\sqrt{V}}{\sqrt{3}H},\label{eqn12}\\
		z:&=\frac{\sqrt{6}}{\phi},\label{eqn13}\\
		u:&=\frac{1}{\sqrt{3}\dot{a}}.\label{eqn14}
	\end{align}
	By using these dynamical variables the cosmic evolution equation (in the last section) can be written in an autonomous system as
	\begin{eqnarray}
		\frac{dx}{dN}&=&-3x-\sqrt{\frac{3}{2}}\left(\frac{V'}{V}\right)y^2+3x(x^2-\mathcal{K} u^2),\label{eq14}\\
		\frac{dy}{dN}&=&3y(x^2-\mathcal{K} u^2)+\sqrt{\frac{3}{2}}\left(\frac{V'}{V}\right)x y,\label{eq15}\\
		\frac{dz}{dN}&=&-xz^2,\label{eq16}\\
		\frac{du}{dN}&=&u(2x^2-y^2)\label{eq17}
	\end{eqnarray}
	together with $N = \ln a$.  Using dynamical variables, the first Friedmann equation (\ref{eq7}) can be rewritten as
	\begin{align}
		\Omega_\mathcal{K}+x^2+y^2=1,\label{eq18}
	\end{align}
	where the density parameter due to the curvature term can be expressed as
	\begin{align}
		\Omega_\mathcal{K}=\frac{\rho_{\mathcal{K}}}{3H^2}= -3\mathcal{K} u^2.\label{eq19}
	\end{align}
It is to be noted that the dynamical variable $u$ is proportional to the comoving Hubble horizon.   The amplitude of the scalar field $\phi$ is inversely related to the variable $z$.   From Eq.(\ref{eq18}) we can clearly conclude about the meaning of the dynamical variables $(\ref{eqn11})$ and $(\ref{eqn12})$: $x^2$ stands for the relative kinetic energy density of the scalar field while $y^2$ stands for its relative potential energy density of $\phi$ and the total relative energy density parameter due to the scalar field can be written as \cite{Bahamonde:2017ize}
	\begin{align}
		\Omega_\phi=\frac{\rho_\phi}{3H^2}=x^2+y^2.\label{eq20}
	\end{align}
	The equation of state parameter ($\omega_\phi$) can be expressed as
	\begin{align}
		\omega_\phi&=\frac{x^2-y^2}{x^2+y^2}
	\end{align}
	and the equation of state parameter corresponding to the combined matter field is given by
	\begin{align}
		\omega_{total}&=x^2-y^2.
	\end{align}

	From (\ref{eq18}), (\ref{eq19}) and (\ref{eq20}), we can conclude that the variables $x$, $y$, and $u$ are not independent, they are dependent by the constraint equation
	\begin{align}
		x^2+y^2-3\mathcal{K}u^2=1.\label{eq23}
	\end{align}
This equation is nothing but the conservation of energy with proper scaling.  In Proposition (\ref{C}), we have shown that (\ref{eq14}), (\ref{eq15}) and (\ref{eq17}) are not independent, in other words Eq.(\ref{eq17}) can be obtained by using Eqs.(\ref{eq14}), (\ref{eq15}) and the constraint equation (\ref{eq23}).  Thus we omit equation (\ref{eq17}) from the autonomous system and by substituting (from Eq.(\ref{eq23}))
	$$
	\mathcal{K}u^2=\frac{1}{3}(x^2+y^2-1)
	$$
	into the right hand side of (\ref{eq14}) and (\ref{eq15}), we get the following modified autonomous system
	\begin{align}
		\frac{dx}{dN}&=-2x-\sqrt{\frac{3}{2}}\left(\frac{V'}{V}\right)y^2-xy^2+2x^3,\label{eqn24}\\ \frac{dy}{dN}&=y+\sqrt{\frac{3}{2}}\left(\frac{V'}{V}\right)xy+2x^2 y-y^3,\label{eqn25}\\ \frac{dz}{dN}&=-xz^2.  \label{eqn26}
	\end{align}
	
Note that the scalar field model corresponds to stiff matter if potential energy (P.E.) is negligible compared to kinetic energy (K.E.), while it corresponds to $\Lambda$CDM if kinetic energy is negligible compared to potential energy.  In between these two extreme cases radiation will occur if K.E.=2 times the P.E.  In that case the dimensionless variables $x$ and $y$ are not independent and the system becomes a $2$D system.  Also as we are interested in late time cosmology so radiation is not taken into account. 	
	
	To determine a suitable form of the autonomous system, we consider several choices of the scalar field potential \cite{PhysRevD.37.3406,Dodelson:2000jtt,Urena-Lopez:2000ewq,Brax:1999gp,Steinhardt:1999nw,Frieman:1995pm, Mishra:2019vnv,Arapoglu, Nandan,Singh:2019enu,soumya}.  But only for power-law \cite{Nandan,Singh:2019enu} $V(\phi)=V_0\phi^{-\mu}$ and the exponential dependence \cite{Mandal:2023nqf,Arapoglu, Nandan,Singh:2019enu, Mishra:2021sza} as $V(\phi)=V_1e^{-\nu\phi}$ where $V_0, V_1>0$ and $\mu,\nu$ are constant parameters, we get the suitable form of the autonomous system.  These two potentials are interesting because we find that scalar-field potential is ``nonlinear'' (i.e., the scalar-field equation of motion is nonlinear); usually it tends to be made from exponential of the scalar field. We have not succeeded in determining the general solution of the
scalar-field equation of motion (\ref{eq9}), but a special solution (cosmologically relevant) can be found. Somehow particularly, we find that this solution dominates at large time and a study of phase space shows that it is an attractive, time-dependent, equilibrium point \cite{Mishra:2021sza}. 
	
	\subsection{Model 1: Power-law potential} 
	
 At first we shall analyze the above dynamical system for power-law potential.  The above system of first order differential equations $(\ref{eqn24}-\ref{eqn26})$ reduces to an autonomous system.  The justification for such a power law form is that it satisfies the sufficiently steep condition namely $\Gamma=\frac{V'' V}{(V')^2}\geq 1$.  As a result, the scalar field rolls down such a potential and favors a common evolutionary path for a wide range of initial conditions \cite{Sahni:2004ai,PhysRevD.37.3406}. The form of power law can be expressed as
	$$
	V(\phi)=V_0\phi^{-\mu}.
	$$
	Substituting this into the right hand side of Equation (\ref{eqn24}) and Equation (\ref{eqn25}) and then the system $(\ref{eqn24}-\ref{eqn26})$ modifies to
	\begin{align}
		\frac{dx}{dN}&=-2x+\frac{\mu}{2}y^2z-xy^2+2x^3,\label{eqn29}\\
		\frac{dy}{dN}&=y-\frac{\mu}{2}xyz+2x^2 y-y^3,\label{eqn30}\\
		\frac{dz}{dN}&=-xz^2.\label{eqn31}
	\end{align}
	
	We have a total of five critical points corresponding to the autonomous system $(\ref{eqn29}-\ref{eqn31})$ out of which four are isolated critical points and another point is a line of the critical point.  The set of critical points and the value of cosmological parameters corresponding to the above autonomous system are shown in Table {\ref{T1}}.  
	
	\begin{table}[h]
		\caption{\label{T1}Table shows the set of critical points, the existence of critical points, and the value of cosmological parameters corresponding to the autonomous system $(\ref{eqn29}-\ref{eqn31})$:}
		\begin{tabular}{|c|c|c c c|c|c|c|c|c|}
			
			\hline
			\begin{tabular}{@{}c@{}}$~~$\\ Critical Points \\$~$\end{tabular}   & Existence& $x$ &$y$&$z$&$\Omega_\mathcal{K}$ &$\Omega_\phi$& $\omega_\phi$ &$\omega_{total}$ & $q$\\ \hline\hline
			\begin{tabular}{@{}c@{}}$~~$\\$ C_1 $\\$~$\end{tabular} & For all $\mu\neq 0$ & $0$ & $1$&$0$&$0$&$1$&$-1$&$-1$&$-1$\\ \hline 
			\begin{tabular}{@{}c@{}}$~$\\$ C_2 $\\$~$\end{tabular}  &For all $\mu\neq 0$  &$0$&$-1$&$0$&$0$&$1$&$-1$&$-1$&$-1$\\ \hline
			\begin{tabular}{@{}c@{}}$~~$\\$ C_3 $\\$~$\end{tabular}  & For all $\mu$ &$1$&$0$&$0$&$0$&$1$&$1$&$1$&$2$\\ \hline 
			\begin{tabular}{@{}c@{}}$~~$\\$ C_4 $\\$~$\end{tabular} & For all $\mu$ & $-1$ &$0$&$0$ &$0$&$1$&$1$&$1$&$2$ \\ \hline 	
			\begin{tabular}{@{}c@{}}$~~$\\$ C_5 $\\$~$\end{tabular} & For all $\mu$ & $0$ &$0$&$z_c$ &$1$&$0$& Not applicable &$0$&$\frac{1}{2}$ \\ \hline 	
		\end{tabular}	
	\end{table}

\subsubsection{ \textbf{Local dynamics around the critical points for power-law potential}}

$\bullet$ The critical points $C_1$ and $C_2$ exist for all $\mu \neq 0$.  Near these points the curvature is negligible.  These solutions are completely dominated by scalar field ($\Omega_{\phi}=1$) which behaves as cosmological constant ($\omega_{total}=-1$).  The cosmic evolution near the points characterizes the de-Sitter expansion ($q=-1$) of the universe.  On the other hand, we note that $|\Omega_\mathcal{K}|\approx 0$.  Then from equation (\ref{eq8}) we have $\dot{H} \approx 0$, i.e., the expansion rate $H$ becomes constant.
	
\begin{figure}
		\includegraphics[width=1\textwidth]{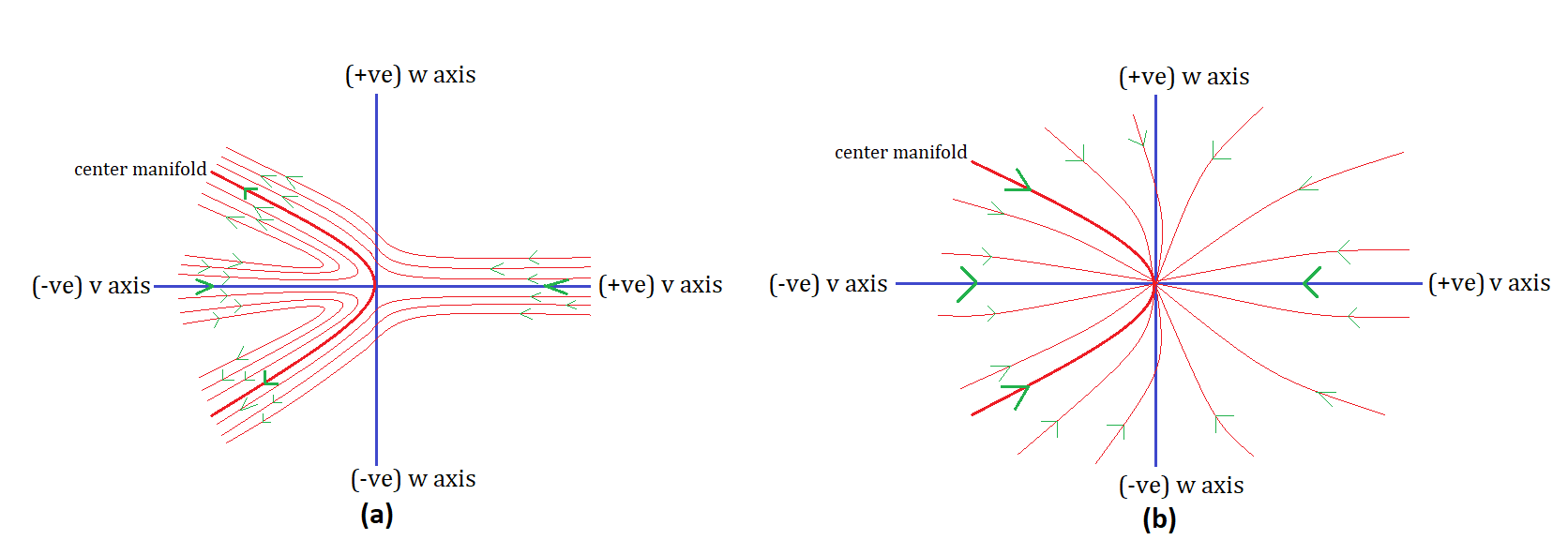}
		\caption{\label{cm}  This figure shows the vector field near the origin corresponding to the critical point $C_1$ for positive and negative values of the parameter $\mu$.  \textbf{(a)} is for $\mu<0$ and \textbf{(b)} is for $\mu>0$.  This figure indicates that for $\mu<0$ the critical point $C_1$ behaves as a saddle node and for $\mu>0$ the critical point $C_1$ behaves as a stable node.}
	\end{figure}

\begin{figure}
			\includegraphics[width=1\textwidth]{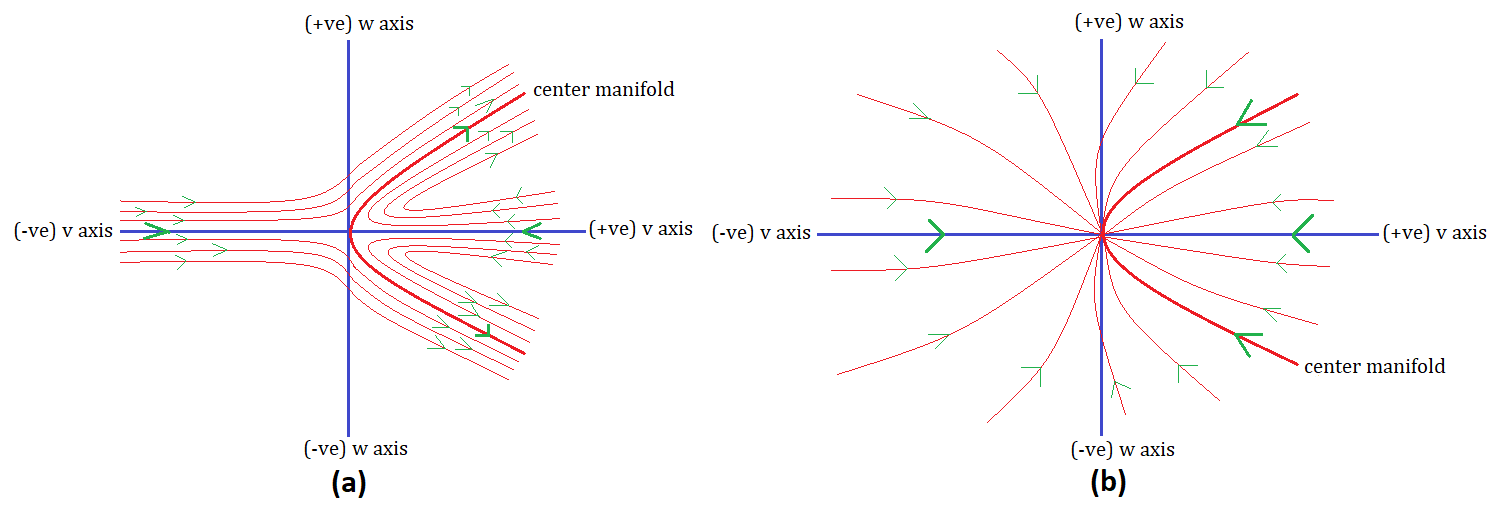}
			\caption{\label{CM_1} This figure shows the vector field near the origin corresponding to the critical point $C_2$ for positive and negative values of the parameter $\mu$.  \textbf{(a)} is for $\mu<0$ and \textbf{(b)} is for $\mu>0$.  This figure indicates that for $\mu<0$ the critical point $C_2$ behaves as a saddle node and for $\mu>0$ the critical point $C_2$ behaves as a stable node.}
		\end{figure}

The eigenvalues of the Jacobian matrix $J(C_1)\left(J(C_2)\right)$ are $\lambda_1=-3$, $\lambda_2=-2$, and $\lambda_3=0$ corresponding to the eigenvectors $\mathbf{v}^{(1)}=[1, 0, 0]^T$, $\mathbf{v}^{(2)}=[0, 1, 0]^T$, and $\mathbf{v}^{(3)}=\left[\frac{\mu}{6}, 0, 1\right]^T$ respectively.  Since the Jacobian matrix has one zero eigenvalue so we use center manifold theory to discuss the stability of the critical point.  To apply center manifold theory \cite{10.5555/102732,soumya, Mishra:2019vnv,Patil:2022uco}, first, we use the shifting transformation $x=X,~ y=Y\pm1,~ z=Z$ to transform the critical point $C_1(C_2)$ into the origin. Next, we use a coordinate transformation $\mathbf{u}=P\mathbf{v}$ where $\mathbf{u}=[X~ Y~Z]^T$, $\mathbf{v}=[u~ v~w]^T$, $P=\left[\mathbf{v}^{(1)}~\mathbf{v}^{(2)}~\mathbf{v}^{(3)}\right]$.  From the calculation (see Proposition (\ref{CPC1}) in the Appendix), the expressions of the center manifold are given by $u=0$ and $v=\mp\frac{\mu^2}{72}w^2+\mathcal{O}(w^4)$ 
	and the flow on the center manifold is determined by $\frac{dw}{dN}=-\frac{\mu}{6}w^3+\mathcal{O}(w^4).$ Notice that the flow on the center manifold depends on the sign of $\mu$.  If $\mu<0$ then $\frac{dw}{dN}>0$ for $w>0$ and $\frac{dw}{dN}<0$ for $w<0$.  So, for $\mu<0$ the origin is a saddle point, i.e., unstable in nature.  Further if $\mu>0$ then $\frac{dw}{dN}<0$ for $w>0$ and $\frac{dw}{dN}>0$ for $w<0$.  Thus for $\mu>0$ the origin is a stable node.  The vector field near the origin for both of the cases are shown in figures (\ref{cm}) and (\ref{CM_1}).  It is to be noted that the new coordinate system $(u,~v,~w)$ is topologically equivalent to the old one, hence the origin in the new coordinate system, i.e., the critical point $C_1(C_2)$ in the old coordinate system $(x,~y,~z)$ is a saddle point for $\mu<0$ and a stable node for $\mu>0$.
	
\begin{figure}
			\includegraphics[width=0.45\textwidth]{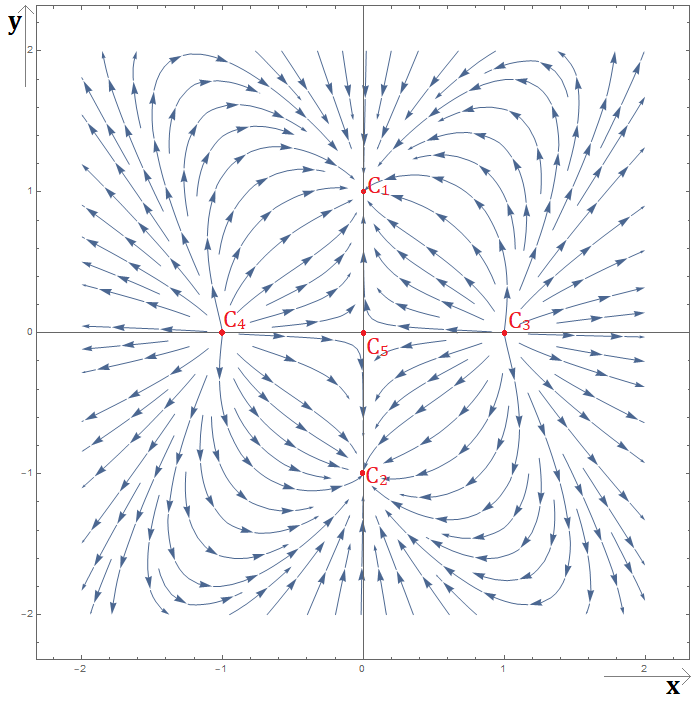}
			\caption{\label{c_12345} Profile of the global analysis in finite phase space for several values of $\mu$.  This figure shows the projection of the vector field on the $xy$ plane corresponding to the autonomous system $(\ref{eqn29}-\ref{eqn31})$.  The horizontal axis represents variable ‘x’ and the vertical axis represents variable ‘y’.  This phase plot indicates that in $xy$ plane the critical points $C_1$ and $C_2$ behave as a stable node, $C_3$ and $C_4$ behave as an unstable node, and the critical point $C_5$ behaves as a saddle node for all $\mu$.}
		\end{figure}

	 If we take the projection of the vector field in the $z=0$ plane, that is, on the $xy$-plane, we can see that the system converges to the critical point $C_1(C_2)$.  Note that the eigenvalues of the Jacobian matrix at the point $C_1(C_2)$ are real and negative, it follows that the critical points $C_1$ and $C_2$ are stable node and asymptotically stable in nature (see figure (\ref{c_12345})).
	
Near $C_1$ and $C_2$ the evolution equation ({\ref{eq9}})  takes the form $\ddot{\phi} + V'(\phi) \approx 0$.  So the scalar field potential behaves as a simple harmonic oscillator and for $\mu >0$ the oscillation amplitude decreases gradually due to the stability of the critical points.  \\

$\bullet$ The critical points $C_3$ and $C_4$ exist for $\mu \in \mathbb{R}$.  The critical points correspond to solutions where the constraint equation ($\ref{eq8}$) is dominated by the kinetic energy of the scalar field with a stiff equation of state $\omega_d=1$.  There exists a decelerating phase of the universe and we also have $|\Omega_\mathcal{K}|\approx 0$ near these critical points.  Then from equation (\ref{eq8}) we have $\dot{H}$ is negative, i.e., the expansion  rate $H$ decreases.

\begin{figure}
			\includegraphics[width=1\textwidth]{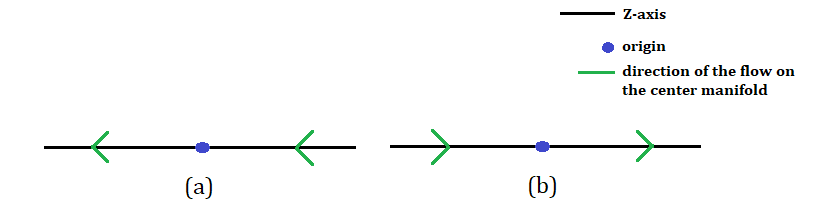}
			\caption{\label{zcentermanifold} This figure shows flow direction on the center manifold corresponding to the critical point $C_3$ and $C_4$.  \textbf{(a)} is for the critical point $C_3$ and \textbf{(b)} is for the critical point $C_4$.  The direction of the flow on the center manifold indicates that both of the critical points are saddle-node in nature.}
		\end{figure}

The eigenvalues of the Jacobian matrix $J(C_3)\left(J(C_4)\right)$ are $4$, $3$, and $0$ corresponding to the eigenvectors $[1, 0, 0]^T$, $[0, 1, 0]^T$ and $\left[0,0,1\right]^T$ respectively.  After applying shifting transformation $x=X+1$, $y=Y$ and $z=Z$ to the autonomous system $(\ref{eqn29}-\ref{eqn31})$ and then by using the center manifold theory, the equation of center manifold can be expressed as $X=0,~ Y=0.$  This implies that the center manifold completely lying on the $Z$ axis and the flow on the center manifold is determined by $\frac{dZ}{dN}=\mp Z^2.$  Notice that $\frac{dZ}{dN}<0$ for both of the cases $Z>0$ or $Z<0$ for the critical point $C_3$, and $\frac{dZ}{dN}>0$ for $Z>0$ or $Z<0$ for the critical point $C_4$.  From this one concludes that the critical point $C_3(C_4)$ is a saddle node, that is, along the $Z$ direction the orbits approach critical point $C_3$ for positive(negative) $Z$ and move away from it for negative(positive) $Z$.  The flow on the center manifold is shown in figure (\ref{zcentermanifold}) which implies that the critical points $C_3$ and $C_4$ behave as a saddle-node.
		
	 Now if we take the projection of the vector field in the $z=0$ plane, that is, on the $xy$-plane, the system diverges from the critical point $C_3(C_4)$.  Note that the eigenvalues are real and positive which implies that the critical point $C_3(C_4)$ is unstable in nature due to the unstable node type instability (see figure (\ref{c_12345})).
		
 If the kinetic energy grows much faster than the radiation, matter, or curvature to prevent the kinetic energy from dominating the universe, the universe continues to contract until its size becomes zero.  So the Big Crunch may happen if the flow of the vector field goes towards  $C_3$ or $C_4$ asymptotically.   As the critical points are saddle in nature, to avoid the Big Crunch we need to fine-tune the initial condition.\\
		
\begin{figure}
			\includegraphics[width=0.5\textwidth]{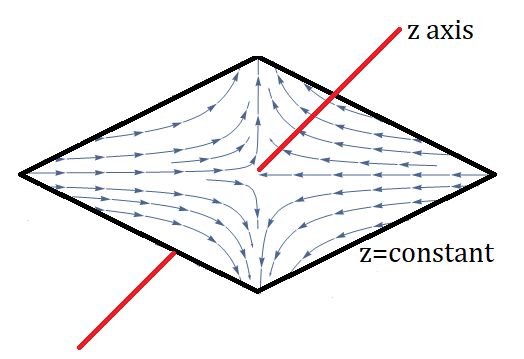}
			\caption{\label{line_of_CPs} This figure shows the projection of the vector field near $C_5$ on $z=const.$ plane.  This phase plot indicates that the line of critical points $C_5$ behaves as a saddle node for all $\mu$.}
		\end{figure}

 $\bullet$ The set of critical points $C_5$ exist for all $\mu \in \mathbb{R}$.  This set of critical points is normally hyperbolic where the stability is confirmed by the signature of the remaining non-vanishing eigenvalues.   The constraint equation (\ref{eq8}) is dominated by spatial curvature near this set of points.  In this case, the kinetic term is subdominant compared to the curvature contribution in eq. (\ref{eq8}).  Then, for $\mathcal{K} >0$, $\dot{H}$ is positive, and the expansion rate $H$ increases. On the other hand, for $\mathcal{K}<0$, $\dot{H}$ is negative, and the expansion rate $H$ decreases.  
Note that the line of critical point $C_5$ arises on the $z$ axis so to determine the vector field near $C_5$ we consider only the first two equations of the autonomous system $(\ref{eqn29}-\ref{eqn31})$ and taking $z=const.$ Due to the presence of one positive and one negative eigenvalue corresponding to the first two equations of the autonomous system, by Hartman-Grobman theorem, we conclude that the line of critical points $C_5$ behave as a saddle node (figure (\ref{line_of_CPs})) for all $\mu$.  So depending on the initial condition there exists a late time (along stable eigen direction) decelerating dust-dominated universe.\\

The eigenvalues of the Jacobian matrix at all critical points corresponding to the autonomous system $(\ref{eqn29}-\ref{eqn31})$ and the nature of critical points are shown in Table {\ref{T2}}.
		\begin{table}[h]
			\caption{\label{T2}Table shows the eigenvalues $(\lambda_1,~\lambda_2,~\lambda_3)$ of the Jacobian matrix corresponding to the autonomous system $(\ref{eqn29}-\ref{eqn31})$ at each critical points and the nature of all critical points:}		 
			\begin{tabular}{|c|c c c|c|}
				\hline	
				\begin{tabular}{@{}c@{}}$~~$\\ Critical Points \\$~$\end{tabular} &$ ~~\lambda_1 $ & $~~\lambda_2$ & $~~\lambda_3$  &  Nature of Critical points \\ \hline\hline
				\begin{tabular}{@{}c@{}}$~~$\\$ C_1 $\\$~$\end{tabular}  & $ -3  $ & $ -2 $& $0$ & Saddle node for $\mu<0$ and stable node for $\mu>0$.\\ \hline
				\begin{tabular}{@{}c@{}}$~~$\\$ C_2 $\\$~$\end{tabular}   &  $-3$   &$ -2 $ & $0$& Saddle node for $\mu<0$ and stable node for $\mu>0$.\\ \hline
				\begin{tabular}{@{}c@{}}$~~$\\$ C_3 $\\$~$\end{tabular}   &  $4$   &$ 3 $ & $0$& Saddle node for all $\mu$ (unstable node in $xy$-plane).\\ \hline   
				\begin{tabular}{@{}c@{}}$~~$\\$ C_4 $\\$~$\end{tabular}   &  $4$   &$ 3 $ & $0$& Saddle node for all $\mu$ (unstable node in $xy$-plane).\\ \hline    
				\begin{tabular}{@{}c@{}}$~~$\\$ C_5 $\\$~$\end{tabular}   &  $-2$   &$ 1 $ & $0$& Saddle node for all $\mu$.\\ \hline                                                                                                                                                                                                                                                                                                                                                                  
			\end{tabular}
		\end{table}
		
		\subsection{Model 2: Exponential potential}
	As before this choice of the potential simplifies the system of equations $(\ref{eqn24}-\ref{eqn26})$	into an autonomous system.  The choice of exponential potential corresponds to an extreme example of quintessence \cite{Sahni:2004ai,PhysRevD.37.3406}.  Exponential potential will remain subdominant if it was so initially.  Further, nucleosynthesis constraints the energy density of the quintessence field to be smaller than the associated background energy density at the early era.  The form of exponential dependence can be expressed as
		$$
		V(\phi)=V_0e^{-\nu \phi}.
		$$
		Substituting this into the right hand side of Eq.(\ref{eqn24}) and Eq.(\ref{eqn25}) and then the system $(\ref{eqn24}-\ref{eqn26})$ modifies to
		\begin{align}
			\frac{dx}{dN}&=-2x+\sqrt{\frac{3}{2}}\nu y^2-xy^2+2x^3,\label{equ30}\\ \frac{dy}{dN}&=y-\sqrt{\frac{3}{2}}\nu xy+2x^2 y-y^3,\label{equ31}\\ \frac{dz}{dN}&=-xz^2. \label{equ32}
		\end{align}

		We can see that the first two equations of the autonomous system $(\ref{equ30}-\ref{equ32})$ are independent of the variable $z$.  So when we are going to determine the critical points, we can see that most of the physical meaningful critical points are lying on the $xy$-plane.  Thus the last equation of the autonomous system, that is, equation (\ref{equ32}) is not interesting to provide the stability criteria of the critical points corresponding to the autonomous system $(\ref{equ30}-\ref{equ32})$.  Further, note that the variable $z$ can not impact physically on this model as all the expressions of the cosmological parameters are independent of $z$.  That's why it is more interesting if we consider $z=0$ and analyze the stability of the critical points in $xy$ plane.  We have a total of seven physically meaningful isolated critical points corresponding to the autonomous system $(\ref{equ30}-\ref{equ31})$.  The set of critical points corresponding to this autonomous system and the value of cosmological parameters are shown in Table {\ref{T3}}.

\subsubsection{\textbf{Local dynamics around the critical points for exponential potential}}

		\begin{table}[h]
			\caption{\label{T3}Table shows the set of critical points, the existence of critical points, and the value of cosmological parameters corresponding to the autonomous system $(\ref{equ30}-\ref{equ31})$:}
			\begin{tabular}{|c|c|c c |c|c|c|c|c|}
				
				\hline
				\begin{tabular}{@{}c@{}}$~~$\\ Critical Points \\$~$\end{tabular}   & Existence& $x$ &$y$&$\Omega_\mathcal{K}$ &$\Omega_\phi$& $\omega_\phi$ &$\omega_{total}$ & $q$\\ \hline\hline
				\begin{tabular}{@{}c@{}}$~~$\\$P_0 $\\$~$\end{tabular} & For all $\nu$ & $0$ &$0$ &$1$&$0$&Not applicable&$0$&$\frac{1}{2}$ \\ \hline 
				\begin{tabular}{@{}c@{}}$~~$\\$P_1 $\\$~$\end{tabular}  & For all $\nu$ &$1$&$0$&$0$&$1$&$1$&$1$&$2$\\ \hline 
				\begin{tabular}{@{}c@{}}$~~$\\$ P_2 $\\$~$\end{tabular} & For all $\nu$ & $-1$ &$0$&$0$ &$1$&$1$&$1$&$2$ \\ \hline 	
				\begin{tabular}{@{}c@{}}$~~$\\$ P_3 $\\$~$\end{tabular}  & For $0\leq\nu^2\leq 6$ &$\frac{\nu}{\sqrt{6}}$&$\sqrt{1-\frac{\nu^2}{6}}$&$0$&$1$&$\frac{\nu^2}{3}-1$&$\frac{\nu^2}{3}-1$&$\frac{1}{2}(\nu^2-2)$\\ \hline 
				\begin{tabular}{@{}c@{}}$~~$\\$ P_4 $\\$~$\end{tabular} & For $0\leq\nu^2\leq 6$ & $\frac{\nu}{\sqrt{6}}$ &$-\sqrt{1-\frac{\nu^2}{6}}$ &$0$&$1$&$\frac{\nu^2}{3}-1$&$\frac{\nu^2}{3}-1$&$\frac{1}{2}(\nu^2-2)$\\ \hline 	
				\begin{tabular}{@{}c@{}}$~~$\\$P_5 $\\$~$\end{tabular} & For all $\nu\neq 0$ & $\frac{1}{\nu}\sqrt{\frac{2}{3}}$ &$\frac{2}{\sqrt{3}\nu}$ &$1-\frac{2}{\nu^2}$&$\frac{2}{\nu^2}$&$-\frac{1}{3}$&$-\frac{2}{3\nu^2}$&$\frac{1}{2}\left(1-\frac{2}{\nu^2}\right)$ \\ \hline 	
				\begin{tabular}{@{}c@{}}$~~$\\$P_6 $\\$~$\end{tabular} & For all $\nu\neq 0$ & $\frac{1}{\nu}\sqrt{\frac{2}{3}}$ &$-\frac{2}{\sqrt{3}\nu}$ &$1-\frac{2}{\nu^2}$&$\frac{2}{\nu^2}$&$-\frac{1}{3}$&$-\frac{2}{3\nu^2}$&$\frac{1}{2}\left(1-\frac{2}{\nu^2}\right)$ \\ \hline 	
			\end{tabular}	
		\end{table}
		
		$\bullet$ The critical point $P_0$ exists for all $\nu$.  For this solution, the kinetic term is subdominant compared to the curvature contribution in eq. (\ref{eq8}).  Then, for $\mathcal{K} >0$, $\dot{H}$ is positive, and the expansion rate $H$ increases. On the other hand, for $\mathcal{K}<0$, $\dot{H}$ is negative, and the expansion rate $H$ decreases.
		
\begin{figure}
			\includegraphics[width=1\textwidth]{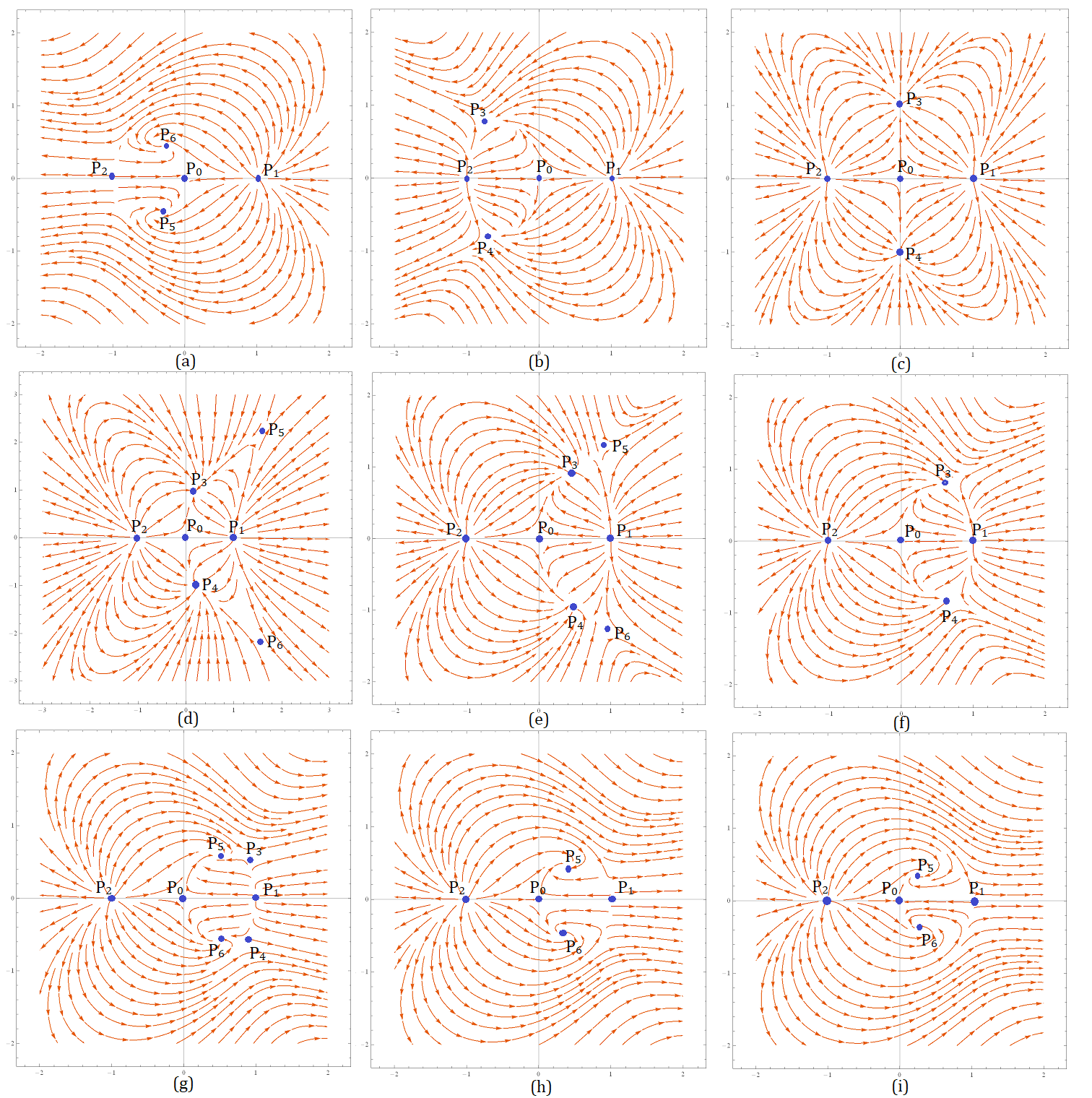}
			\caption{\label{P_012345} Profile of the global analysis in finite phase space for several values of $\nu$.  The horizontal axis represents variable $\lq$x' and the vertical axis represents variable $\lq$y'. (a) corresponds to $\nu=-\sqrt{6}$ : $P_0$ and $P_2$ are saddle node, $P_1$ is an unstable node, and $P_5$ and $P_6$ are spiral sink, (b) corresponds to $\nu=-\sqrt{2}$ : $P_0$, $P_3$ and $P_4$ are saddle node, $P_1$ and $P_2$ are unstable node, (c) corresponds to $\nu=0$ : $P_0$ is a saddle node, $P_1$ and $P_2$ are unstable node, and $P_3$ and $P_4$ are stable node, (d) corresponds to $\nu=\frac{1}{2}$ : where $P_0$ is a saddle node, $P_1$ and $P_2$ are unstable node, $P_3$ and $P_4$ are stable node, and $P_5$ and $P_6$ are saddle node, (e) corresponds to $\nu=1$ : $P_0$ is a saddle node, $P_1$ and $P_2$ are unstable node, $P_3$ and $P_4$ are stable node, and $P_5$ and $P_6$ are saddle node, (f) corresponds to $\nu=\sqrt{2}$ : $P_0$, $P_3$ and $P_4$ are saddle node, $P_1$ and $P_2$ are unstable node, (g) corresponds to $\nu=2$ : $P_0$, $P_3$ and $P_4$ are saddle node, $P_1$ and $P_2$ are unstable node, and $P_5$ and $P_6$ are spiral sink, (h) corresponds to $\nu=\sqrt{6}$ : $P_0$ and $P_1$ are saddle node, $P_2$ is an unstable node, and $P_5$ and $P_6$ are spiral sink, and (i) corresponds to $\nu=3$ : $P_0$ and $P_1$ are saddle node, $P_2$ is an unstable node, and $P_5$ and $P_6$ are spiral sink. }
		\end{figure}

	The eigenvalues of the Jacobian matrix $J(P_0)$ are $\lambda_1=-2$, and $\lambda_2=1$.  Since $J(P_0)$ has no zero eigenvalue, the fixed point $P_0$ is hyperbolic.  Further, note that $\lambda_1<0$ and $\lambda_2>0$, thus by using the Hartman-Grobman theorem, we conclude that the critical point $P_0$ behaves as a saddle node (see figure (\ref{P_012345})).
			 Depending on the initial condition there exists a late time (along stable eigen direction) decelerating dust-dominated universe.\\

		$\bullet$ Similar to the case of $C_3$ and $C_4$, the critical point $P_1$ and $P_2$ exist for all $\nu$ and correspond to the solutions where the constraint equation ($\ref{eq8}$) is dominated by the kinetic energy of the scalar field with a stiff equation of state $\omega_d=1$.  There exists a decelerating phase of the universe and we also have $|\Omega_\mathcal{K}|\approx 0$ near these critical points.  Then from equation (\ref{eq8}) we have $\dot{H}$ is negative, i.e. the expansion rate $H$ decreases.
		
	 The eigenvalues of the Jacobian matrix $J(P_1)$ are $\lambda_1=4$, and $\lambda_2=3-\sqrt{\frac{3}{2}}\nu$. Notice that the eigenvalue $\lambda_1$ is always positive which implies that the critical point $P_1$ is always unstable in nature but the type of instability is determined by the nature of the eigenvalue $\lambda_2$. The critical point $P_1$ is hyperbolic for $\nu\neq \sqrt{6}$ and nonhyperbolic for $\nu=\sqrt{6}$.  For $\nu<\sqrt{6}$, the eigenvalue $\lambda_2$ is positive which implies that the critical point $P_1$ is an unstable node.  For $\nu>\sqrt{6}$, we have $\lambda_2<0$ and consequently the critical point $P_1$ is a saddle node.  To determine the stability of $P_1$ for $\nu=\sqrt{6}$, we use center manifold theory which we have already applied to discuss the stability of the critical point $C_1$.  To apply this theory, first, we use a shifting transformation $(x=X+1,y=Y)$ to transform the fixed point $P_1$ into the origin.  Then using the similar arguments that we have used to determine the stability analysis of $C_1$, we obtained that the expression of the center manifold is $X=-\frac{1}{2}Y^2+\mathcal{O}(Y^3)$ and the flow on the center manifold is determined by the equation $\frac{dY}{dN}=-\frac{3}{2}Y^3+\mathcal{O}(Y^4)$.  Note that $\frac{dY}{dN}<0$ while $Y>0$ and $\frac{dY}{dN}>0$ for $Y<0$.  This implies that the flow on the center manifold is stable but the origin is a saddle node.  As the transformed system $(X,Y)$ is topologically equivalent to the old one, the critical point $P_1$ also exhibits saddle-node type instability for $\nu=\sqrt{6}$.  Hence, the critical point $P_1$ is a saddle-node for $\nu\geq \sqrt{6}$ and unstable node for $\nu<\sqrt{6}$ (see figure (\ref{P_012345})).		
		
		 The eigenvalues of the Jacobian matrix $J(P_2)$ are $\lambda_1=4$, and $\lambda_2=3+\sqrt{\frac{3}{2}}\nu$. Notice that the eigenvalue $\lambda_1$ is always positive which implies that the critical point $P_2$ is always unstable in nature but the type of instability is determined by the nature of the eigenvalue $\lambda_2$. The critical point $P_2$ is hyperbolic for $\nu\neq -\sqrt{6}$ and nonhyperbolic for $\nu=-\sqrt{6}$.  For $\nu>-\sqrt{6}$, the eigenvalue $\lambda_2$ is positive which implies that the critical point $P_2$ is an unstable node.  For $\nu<-\sqrt{6}$, we have $\lambda_2<0$ and consequently the critical point $P_2$ is a saddle node.  To determine the stability of $P_2$ for $\nu=\sqrt{6}$, we use center manifold theory and to apply this theory, first, we use a shifting transformation $(x=X-1,y=Y)$ to transform the fixed point $P_2$ into the origin.  Then the expression of the center manifold is given by $X=\frac{1}{2}Y^2+\mathcal{O}(Y^3)$ and the flow on the center manifold is determined by the equation $\frac{dY}{dN}=-\frac{3}{2}Y^3+\mathcal{O}(Y^4)$.  Note that $\frac{dY}{dN}<0$ while $Y>0$ and $\frac{dY}{dN}>0$ for $Y<0$.  This implies that the flow on the center manifold is stable but the origin is a saddle node.  Hence, the critical point $P_2$ is a saddle-node for $\nu\leq -\sqrt{6}$ and unstable node for $\nu>-\sqrt{6}$ (see figure (\ref{P_012345})).	
		
		 So the Big Crunch can be avoided for $-\sqrt{6} <\nu < \sqrt{6}$ due to the instability of the critical points.  One may note that depending on the initial condition there exists a late time (along stable eigen direction) kinetic dominated decelerating universe.\\

		$\bullet$ The fixed point $P_3$ and $P_4$ exist for  $\nu^2\leq 6$.  Varying $\nu$ we get a non-isolated set of critical points which are completely dominated by kinetic energy ($\Omega_{\phi}=1$ and $\dot{\phi} \neq 0$).  The DE can represent either quintessence or any other exotic type fluid depending on the parameter $\nu$.  Specially, the depletion $\nu ^2 <2 ~ (\nu \neq 0)$ implies that the scalar field behaves as quintessence like fluid and there exists an accelerated universe (since for this case, $-1<\omega_{total}<-\frac{1}{3}, q<0$) near these critical points whereas the scalar field DE behaves as dust for $\nu ^2 =3 ~(\omega_{\phi}=0)$ and in this case the solutions insinuate dust-dominated decelerating phase ($\omega_{total}=0,~q=\frac{1}{2},~\Omega_{\mathcal{K}}=0,~\Omega_{\phi}=1$) of the cosmic evolution.
				
		The eigenvalues of the Jacobian matrix $J(P_3)$ are $\lambda_1=-2+\nu^2$, and $\lambda_2=\frac{1}{2}(\nu^2-6)$. The critical point $P_3$ is hyperbolic for $\nu\neq \pm \sqrt{2},\pm \sqrt{6}$ because $\lambda_1=0$ for $\nu=\pm \sqrt{2}$ and $\lambda_2=0$ for $\nu=\pm \sqrt{6}$.  Consequently, the critical point $P_3$ is nonhyperbolic if $\nu=\pm \sqrt{2}$ or $\nu=\pm \sqrt{6}$.
						
			\textbf{Hyperbolic case} $\left(\nu\neq \pm \sqrt{2},\pm \sqrt{6}\right)$: As the region of existence of the critical point is $-\sqrt{6}\leq \nu \leq \sqrt{6}$, to determine the stability of this critical point in this case, we consider three intervals (i) $-\sqrt{6}<\nu<-\sqrt{2}$, (ii) $-\sqrt{2}<\nu<\sqrt{2}$, and (iii) $\sqrt{2}<\nu<\sqrt{6}$.  For $\nu\in (-\sqrt{6},-\sqrt{2})$, we have $\lambda_1>0$ and $\lambda_2<0$, it follows that the critical point $P_3$ is a saddle node.  For $\nu\in (-\sqrt{2},\sqrt{2})$, we can see that $\lambda_1,\lambda_2<0$ and this implies that $P_3$ is a stable node and asymptotically stable in nature.  Lastly, for $\nu\in (\sqrt{2},\sqrt{6})$, we have $\lambda_1>0$ and $\lambda_2<0$, thus in this interval of $\nu$ the critical point $P_3$ behaves as a saddle node (see figure (\ref{P_012345})).
			
			\textbf{Nonhyperbolic case} $\left(\nu= \sqrt{2}~\text{or}~\nu= -\sqrt{2}~\text{or}~\nu=\sqrt{6}~\text{or}~\nu=-\sqrt{6}\right)$: 
			
			We use the center manifold theory to determine the stability of $P_3$ for this case.  To apply this theory, first, we use a shifting transformation $\left(x=X+\frac{\nu}{\sqrt{6}},y=Y+\sqrt{1-\frac{\nu^2}{6}}\right)$ to transform the critical point $P_3$ into the origin.  Next, we use a coordinate transformation $\mathbf{u}=P\mathbf{v}$ where $\mathbf{u}=[X~ Y]^T$, $\mathbf{v}=[U~ V]^T$, $P=[\mathbf{v_1}~\mathbf{v_2}]$ and $\mathbf{v_1}$, and $\mathbf{v_2}$ are the eigenvectors corresponding to the eigenvalues $\lambda_1$, and $\lambda_2$ respectively.  For $\nu=\sqrt{2}$, the equations of center manifold is given by $U=\sqrt{\frac{3}{2}}V^2+\mathcal{O}(V^3)$, and the flow on the center manifold is determined by $\frac{dV}{dN}= 4\sqrt{6}V^2+\mathcal{O}(V^3)$.  Note that $\frac{dV}{dN}>0$ for $V>0$ or $V<0$ which implies that the origin is a  saddle node (see figure (\ref{P_012345})).  For $\nu=-\sqrt{2}$, the expression of the center manifold and the flow on the center manifold are the same as of $\nu=\sqrt{2}$ case and consequently, the origin exhibits a similar type of stability in this case also.  For $\nu=\sqrt{6}$ the critical point $P_3$ converts into $P_1$ and for $\nu=-\sqrt{6}$, $P_3$ converts into $P_2$ and the stability analysis of these critical points already discussed.\\
				
	 The eigenvalues of the Jacobian matrix $J(P_4)$ are $\lambda_1=-2+\nu^2$, and $\lambda_2=\frac{1}{2}(\nu^2-6)$. The critical point $P_4$ is hyperbolic for $\nu\neq \pm \sqrt{2},\pm \sqrt{6}$ because $\lambda_1=0$ for $\nu=\pm \sqrt{2}$ and $\lambda_2=0$ for $\nu=\pm \sqrt{6}$.  Consequently, the critical point $P_4$ is nonhyperbolic if $\nu=\pm \sqrt{2}$ or $\nu=\pm \sqrt{6}$.			
			
			\textbf{Hyperbolic case} $\left(\nu\neq \pm \sqrt{2},\pm \sqrt{6}\right)$: As the region of existence of the critical point is $-\sqrt{6}\leq \nu \leq \sqrt{6}$, to determine the stability of this critical point in this case, we consider three intervals (i) $-\sqrt{6}<\nu<-\sqrt{2}$, (ii) $-\sqrt{2}<\nu<\sqrt{2}$, and (iii) $\sqrt{2}<\nu<\sqrt{6}$.  For $\nu\in (-\sqrt{6},-\sqrt{2})$, we have $\lambda_1>0$ and $\lambda_2<0$ which implies that the critical point $P_4$ is a saddle node.  Now while $\nu\in (-\sqrt{2},\sqrt{2})$, we can see that $\lambda_1,\lambda_2<0$ and this implies that $P_4$ is a stable node and asymptotically stable in nature.  Lastly, for $\nu\in (\sqrt{2},\sqrt{6})$, we have $\lambda_1>0$ and $\lambda_2<0$ which means that the critical point $P_4$ is a saddle node (see figure (\ref{P_012345})).
			
			\textbf{Nonhyperbolic case} $\left(\nu= \sqrt{2}~\text{or}~\nu= -\sqrt{2}~\text{or}~\nu=\sqrt{6}~\text{or}~\nu=-\sqrt{6}\right)$: 
			
			We use the center manifold theory to determine the stability of $P_4$ for this case.  To apply this theory, first, we use a shifting transformation $\left(x=X+\frac{\nu}{\sqrt{6}},y=Y-\sqrt{1-\frac{\nu^2}{6}}\right)$ to transform the critical point $P_4$ into the origin.  Next, we use a coordinate transformation $\mathbf{u}=P\mathbf{v}$ where $\mathbf{u}=[X~ Y]^T$, $\mathbf{v}=[U~ V]^T$, $P=[\mathbf{v_1}~\mathbf{v_2}]$ and $\mathbf{v_1}$, and $\mathbf{v_2}$ are the eigenvectors corresponding to the eigenvalues $\lambda_1$, and $\lambda_2$ respectively.  For $\nu=\sqrt{2}$, the equations of center manifold is given by $U=-\sqrt{\frac{3}{2}}V^2+\mathcal{O}(V^3)$, and the flow on the center manifold is determined by $\frac{dV}{dN}=- 4\sqrt{6}V^2+\mathcal{O}(V^3)$.  Note that $\frac{dV}{dN}<0$ for $V>0$ or $V<0$ which concludes that the origin is a saddle node (see figure (\ref{P_012345})).  For $\nu=-\sqrt{2}$, the expression of the center manifold and the flow on the center manifold are the same as of $\nu=\sqrt{2}$ case and consequently, the origin exhibits a similar type of stability in this case also.  For $\nu=\sqrt{6}$ the critical point $P_4$ converts into $P_1$ and for $\nu=-\sqrt{6}$, $P_4$ converts into $P_2$ and the stability analysis of these critical points already discussed.
		
	Therefore, one can conclude that the points $P_3$ and $P_4$ correspond to kinetic dominated solutions that behave as quintessence like fluid ($\nu^2<2$) and there exists an accelerated universe at late times, although cosmological coincidence problem can not be alleviated by these points.  On the other hand, late time decelerated solution ($2\leq\nu^2\leq 6$) depends on the initial condition of the autonomous system of model 2. 		
		
	$\bullet$ Scaling solutions represented by the critical points $P_5$ and $P_6$ exist for all $\nu \neq 0$.   Varying $\nu$ we get a non-isolated set of critical points where the nature (hyperbolic or normally-hyperbolic) of points depends on the values of $\nu$.  Near these critical points, scalar field DE behaves as a quintessence boundary ($\omega_{\phi}=-\frac{1}{3}$).  As $0\leq \Omega_{\phi}\leq 1$ when $\mathcal{K}<0$ (from eq.(\ref{eq23})), so for cosmological viability the parameter  $\nu$ must satisfy $\nu^2 \geq 2$.   In this case, there exists always a decelerating universe.    On the other hand, for $\mathcal{K}>0$  there exists an accelerating universe when $\nu^2<2$.   The cosmic coincidence problem can be alleviated by these points depending on the values of $\nu$.
				
		 The eigenvalues of the Jacobian matrices $J(P_5)$ and $J(P_6)$ are same and those are $\lambda_1=\frac{-\nu^2+\sqrt{(8-3\nu^2)\nu^2}}{\nu^2} $, and $\lambda_2=\frac{-\nu^2-\sqrt{(8-3\nu^2)\nu^2}}{\nu^2}$. Thus the critical points $P_5$ and $P_6$ are hyperbolic for $\nu\neq \pm\sqrt{2}$ and nonhyperbolic for $\nu=\pm \sqrt{2}$.  For $\nu^2<2$, we can observe that $\lambda_1>0$ and $\lambda_2<0$ which concludes that $P_5$ and $P_6$ saddle point and unstable in nature.  For $2<\nu^2<\frac{8}{3}$, we can see that $\lambda_1,\lambda_2<0$ which implies that the critical points $P_5$ and $P_6$ are asymptotically stable in nature due to the stable node type stability.  For $\nu^2>\frac{8}{3}$, the eigenvalues are complex conjugate to each other with negative real part which implies that the critical points $P_5$ and $P_6$ are spiral sink and asymptotically stable in nature (see figure (\ref{P_012345})).  For $\nu=\pm \sqrt{2}$, the coordinate of $P_5(P_6)$ is the same as of $P_3(P_4)$ for $\nu=\pm \sqrt{2}$ and the stability analysis corresponding to these two cases already discussed.  From the discussion, we conclude that both of the critical points $P_5$ and $P_6$ behave as a saddle node  for $\nu=\pm \sqrt{2}$.
		The eigenvalues of the Jacobian matrix at all critical points corresponding to the autonomous system $(\ref{equ30}-\ref{equ31})$ and the nature of critical points are shown in Table \ref{TEP}.

Therefore, the kinetic dominated solution at $\nu^2=2$ is saddle-node in nature.   For sufficiently flat potential $\nu^2\leq 2$ one may get an accelerating universe at late time depending on the initial condition of the model 2.   Depending on some parameter restrictions, the decelerating universe is the late-time attractor described by these points. \par

		 If we consider our analysis in the three-dimensional coordinate system, that is, in $(x,y,z)$ coordinate system, then we can observe that for all possible cases, the critical points are nonhyperbolic in nature.  To determine the stability of the critical points by applying the center manifold theory, we conclude that for all critical points, the center manifolds are lying on the $z$-axis and the flows on the center manifold are unstable in nature.

		\begin{table}[h]
			\caption{\label{TEP}{Table shows the eigenvalues $(\lambda_1,~\lambda_2)$ of the Jacobian matrix corresponding to the autonomous system $(\ref{equ30}-\ref{equ31})$ at each critical points and the nature of all critical points:}}		 
			\begin{tabular}{|c|c c|c|}
				\hline	
				\begin{tabular}{@{}c@{}}$~~$\\ Critical points \\$~$\end{tabular} &$ \lambda_1 $ & $\lambda_2$  &  Nature of Critical points \\ \hline\hline
				\begin{tabular}{@{}c@{}}$~~$\\$ P_0 $\\$~$\end{tabular}   &  $-2$   &$ 1 $  & Saddle node for all $\nu$.\\ \hline   
				\begin{tabular}{@{}c@{}}$~~$\\$ P_1 $\\$~$\end{tabular}   & $ 4  $ & $ 3-\sqrt{\frac{3}{2}}\nu $ & Saddle node for $\nu\geq \sqrt{6}$ and unstable node for $\nu<\sqrt{6}$.\\ \hline
				\begin{tabular}{@{}c@{}}$~~$\\$ P_2 $\\$~$\end{tabular}  &  $4$   &$ 3+\sqrt{\frac{3}{2}}\nu $ & Saddle node for $\nu\leq -\sqrt{6}$ and unstable node for $\nu>-\sqrt{6}$.\\ \hline
				\begin{tabular}{@{}c@{}}$~~$\\$ P_3 $\\$~$\end{tabular}  &  $-2+\nu^2$   &$ \frac{1}{2}(\nu^2-6)$ & Stable node for $\nu^2<2$ and saddle-node for $2\leq\nu^2\leq 6$.\\ \hline
				\begin{tabular}{@{}c@{}}$~~$\\$ P_4 $\\$~$\end{tabular}  &  $-2+\nu^2$   &$ \frac{1}{2}(\nu^2-6)$ & Stable node for $\nu^2<2$ and saddle-node for $2\leq\nu^2\leq 6$. \\ \hline
				\begin{tabular}{@{}c@{}}$~~$\\$ P_5 $\\$~$\end{tabular}  &  $\frac{-\nu^2+\sqrt{(8-3\nu^2)\nu^2}}{\nu^2}$   &$\frac{-\nu^2-\sqrt{(8-3\nu^2)\nu^2}}{\nu^2}$ & \begin{tabular}{@{}c@{}}$~~$\\Saddle nodet for $\nu^2\leq 2$, stable node for $2<\nu^2\leq \frac{8}{3}$,\\ and spiral sink while $\nu^2>\frac{8}{3}$.\\$~$\end{tabular}\\ \hline
				\begin{tabular}{@{}c@{}}$~~$\\$ P_6 $\\$~$\end{tabular}  &  $\frac{-\nu^2+\sqrt{(8-3\nu^2)\nu^2}}{\nu^2}$   &$\frac{-\nu^2-\sqrt{(8-3\nu^2)\nu^2}}{\nu^2}$ & 	\begin{tabular}{@{}c@{}}$~~$\\Saddle node for $\nu^2\leq 2$, stable node for $2<\nu^2\leq \frac{8}{3}$,\\ and spiral sink while $\nu^2>\frac{8}{3}$.\\$~$\end{tabular}\\ \hline                       
			\end{tabular}
		\end{table}

		\section{Compactification and the dynamics around the critical points at infinity \label{Infinity} }
		
		The idea to compactify the space $\mathbb{R}^n$ by the addition of points at infinity and to map them into finite points is frequently used in the setting of two-dimensional differential equations \cite{Bouhmadi-Lopez:2016dzw,GINGOLD2004284,Leon:2020ovw}.  An early study of analyzing the global behavior of a planar dynamical system via compactification was carried out by Bendixson using the stereographic projection of the sphere (often called Bendixson sphere) onto the plane.  However, this stereographic projection has the drawback of obliterating the ``different directions at infinity'' \cite{ELIAS2006305}.  French mathematician Henri Poincar\'{e} overcame this difficulty by projecting $\mathbb{R}^2$ onto the Poincar\'{e} hemisphere through its center.
		
		In this compactification scheme, we draw straight lines starting from the center of the unit sphere $S^2=\{(X, Y, Z)\in \mathbb{R}^3\mid X^2+Y^2+Z^2=1\}$ to $xy$-plane which is tangent to $S^2$ at either the north or south pole.    It is to be noted that each straight line meets the unit sphere once and the $xy$ plane once.  Let a straight line intersects the sphere at $(X_1, Y_1, Z_1)$ and the $xy$-plane at $(x_1,y_1)$, then our projective transformation is realized by the transformation
		\begin{align}
			x_1=\frac{X_1}{Z_1},~y_1=\frac{Y_1}{Z_1}. \label{p1}
		\end{align}
		
		So an arbitrary point $(x,y)$ on the $xy$-plane can be expressed uniquely by a point $(X,Y,Z)$ on the upper/lower half of the sphere as (\ref{p1}) \cite{ELIAS2006305}.  This scheme has the advantage that the critical points of different directions at infinity are spread out along the equator of the sphere.  
		
		Given an autonomous system of differential equations on $\mathbb{R}^2$ 
		\begin{align}
			&x'=P(x,y), \label{p2}	\\ &  y'=Q(x,y) \label{p3}
		\end{align}
		where $P$ and $Q$ are polynomial functions of $x$ and $y$.   We can write (\ref{p2}) and (\ref{p3}) in the form of a single differential equation 
		$$
		\frac{dy}{dx}=\frac{Q(x,y)}{P(x,y)}
		$$
		which yields
		\begin{align}
			Q(x,y)dx-P(x,y)dy=0. \label{p4}
		\end{align}
		
		Now for any point $(x,y)$, using (\ref{p1}), we obtain
		\begin{align}
			dx=\frac{ZdX-XdZ}{Z^2},~dy=\frac{ZdY-YdZ}{Z^2}.\label{p5}
		\end{align}
		Hence, the differential equation (\ref{p4}) can be represented as
		\begin{align}
			Q(ZdX-XdZ)-P(ZdY-YdZ)=0 \label{p5a}
		\end{align}	
		where
		$$
		P=P(x,y)=P\left(X/Z,Y/Z\right)
		$$	
		and	
		$$
		Q=Q(x,y)=Q\left(X/Z,Y/Z\right).
		$$

		Let $m$ be the maximum degree of the terms in $P$ and $Q$.  We multiply both side of the equation (\ref{p5a}) by $Z^m$  to eliminate $Z$ from the denominator and obtain	
		\begin{align}
			ZQ^*dX-ZP^*dY+(YP^*-XQ^*)dZ=0\label{p6}
		\end{align}
		where
		$$
		P^*(X,Y,Z)=Z^mP(X/Z,Y/Z)
		$$	
		and
		$$
		Q^*(X,Y,Z)=Z^mQ(X/Z,Y/Z)
		$$	
		are polynomials in $(X,Y,Z)$. \\

		The equator of $S^2$ can be expressed by $\{(X,Y,0 ) | X^2+Y^2=1\}$.  So, the critical points of (\ref{p6}) on the equator of $S^2$ where $Z = 0$ are given by the equation	
		\begin{align}
			XQ^*-YP^*=0.  \label{p7}
		\end{align}
		On the equator of the Poincar\'{e} sphere we can derive
		\begin{eqnarray}
			XQ^*-YP^* &= & XQ_m(X,Y)-YP_m(X,Y) \\
			&=& \mathcal{S}~~ \mbox{(say)} \nonumber,
		\end{eqnarray}
		where $P_m$ and $Q_m$ are homogeneous $m^{th}$ degree polynomials in $x$ and $y$ of the system $(\ref{p2})-(\ref{p3})$.  So the critical points at infinity are the set 
		$\{(X, Y, 0)\mid X^2 + Y^2 = 1 ~\mbox{and}~ \mathcal{S}=0 \}$.  Here we do not consider the Eq.(\ref{eqn31}) of the autonomous system $(\ref{eqn29}-\ref{eqn31})$ and Eq.(\ref{equ32}) of the autonomous system $(\ref{equ30}-\ref{equ32})$.  It is to be noted that we get vanishing center manifold equation and vanishing flow near the point at infinity for both of the models for the two equations (\ref{eqn31}) and (\ref{equ32}).  So it is convenient to study the characteristics of the vector field on $xy$ plane for the autonomous systems $(\ref{eqn29}-\ref{eqn31})$ and $(\ref{equ30}-\ref{equ32})$.  Then both of the above autonomous systems can be converted to the following two-dimensional system
		\begin{align}
		\frac{dx}{dN}&=-2x-xy^2+2x^3\label{eqn74},\\ \frac{dy}{dN}&=y+2x^2 y-y^3.\label{eqn75}
		\end{align}
		For the autonomous system $(\ref{eqn74}-\ref{eqn75})$, we obtain the equation (\ref{p7}) as 
		\begin{align}
			XQ_3(X,Y)-YP_3(X,Y) =  X(2X^2Y-Y^3)-Y(-XY^2+2X^3)=0. \label{p8a}
		\end{align}
		Hence, the critical points on the Poincar\'{e} sphere are $(X_c,Y_c,0)$ with $X_c^2+Y_c^2=1$.\\

\begin{figure}
			\begin{subfigure}[b]{0.45\textwidth}
				\centering
				\includegraphics[width=\textwidth]{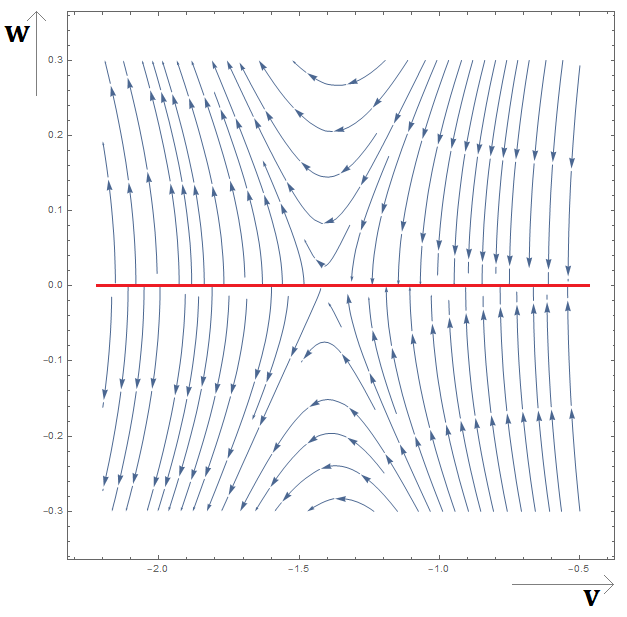}
				\caption{}
				\label{PSa1}
			\end{subfigure}
			\hfill
			\begin{subfigure}[b]{0.45\textwidth}
				\centering
				\includegraphics[width=\textwidth]{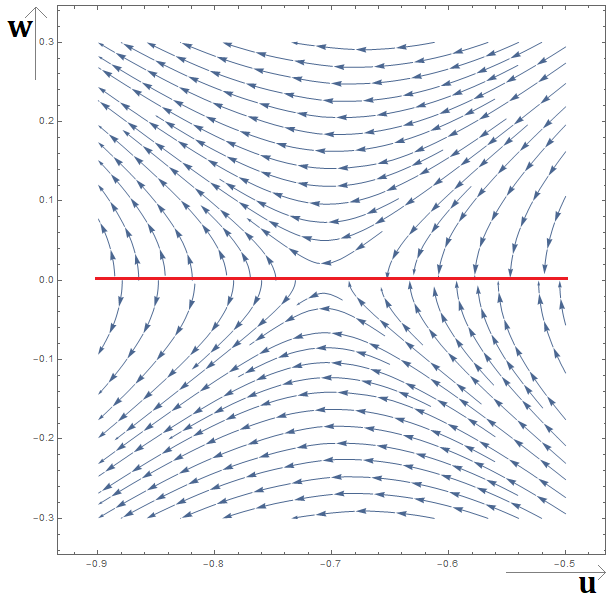}
				\caption{}
				\label{PSb1}
			\end{subfigure}
			\caption{Vector fields near non-isolated critical points except $(0,\pm 1,0)$ are shown in figure (a), and vector fields near non-isolated critical points except $(\pm 1, 0,0)$ are shown in figure (b) on the equator of the Poincar\'{e} sphere.}
			\label{PS1}
		\end{figure}

		It is to be noted that the critical points are non-isolated in nature and the flow in a neighborhood of any critical point $(X_c,Y_c,0)$ on the equator of the Poincar\'{e} sphere $S^2$, except the points $(0, \pm 1, 0 )$, is topologically equivalent to the flow defined by the following system
		
		\begin{align}
			\begin{split}
				\pm \frac{dv}{dN}&=vw^mP\left(\frac{1}{w},\frac{v}{w}\right)-w^m Q\left(\frac{1}{w},\frac{v}{w}\right)	\\ 
				\pm 	\frac{dw}{dN}&=w^{m+1}P\left(\frac{1}{w},\frac{v}{w}\right)
				\label{poin1}
			\end{split}
		\end{align}
		where \\
		\begin{eqnarray*}
			P \left(\frac{1}{w},\frac{v}{w}\right) &=& -\frac{2}{w}-\frac{v^2}{w^3}+\frac{2}{w^3},\\
			Q\left(\frac{1}{w},\frac{v}{w}\right) &=& \frac{v}{w}+2\frac{v}{w^3}-\frac{v^3}{w^3}.
		\end{eqnarray*}
		or equivalently,
		
		\begin{eqnarray}
		\frac{dv}{dN}&=3vw^2,  \label{PCS1} \\   
		\frac{dw}{dN}&=-2w+2w^3+v^2 w.
		\end{eqnarray}
		
		by considering negative signs.  Note that $(v_c,0)$ is non-isolated critical points for the above system (\ref{PCS1}).  The eigenvalues of the Jacobian matrix at $(v_c,0)$ are $0$ and $-2+v_c^2$.  For $|v_c|\leq 1$, we get so $-2+v_c^2<0$ which follows that the vector field near the critical point $(v_c,0)$ is stable along $w$ axis.  On the other hand, for $w\neq 0$ and $v>0$, we have form equation (\ref{PCS1}) that $\frac{dv}{dN}>0$ and for $v<0$, we have $\frac{dv}{dN}<0$.  One may note that the flow comes towards the line of critical points rapidly compare to it goes away from the critical line.  So we can conclude that except the points $(0,\pm 1, 0)$, the other non-isolated points on the equator of the Poincar\'{e} sphere are  saddle node (see figure (\ref{PSa1})) in nature and cosmological bounce may happen near the points at infinity (see section \ref{Bounce}).

		Similarly, the flow in a neighborhood of any critical point $(X_c,Y_c,0)$ on the equator of the Poincar\'{e} sphere $S^2$, except the points $(\pm 1, 0 ,0)$, is topologically equivalent to the flow defined by the following system
		\begin{align}
			\begin{split} 
		\pm	\frac{du}{dN}&=uw^mQ\left(\frac{u}{w},\frac{1}{w}\right)-w^mP\left(\frac{u}{w},\frac{1}{w}\right)\\
		\pm	\frac{dw}{dN}&=w^{m+1}Q\left(\frac{u}{w},\frac{1}{w}\right)
				\label{poin2}
			\end{split}
		\end{align}
		where \\
		\begin{eqnarray*}
			P \left(\frac{u}{w},\frac{1}{w}\right) &=&-2\frac{u}{w}-\frac{u}{w^3}+2\frac{u^3}{w^3} ,\\
			Q\left(\frac{u}{w},\frac{1}{w}\right) &=&\frac{1}{w}+2\frac{u^2}{w^3}-\frac{1}{w^3} .
		\end{eqnarray*}
		or equivalently,
		
		\begin{eqnarray}
		\frac{du}{dN}&=3uw^2, \label{PCS2} \\    
		\frac{dw}{dN}&=-w+2u^2w+w^3. 
		\end{eqnarray}
		
		by considering positive signs. Notice that $(u_c,0)$ is non-isolated critical points for the above system (\ref{PCS2}).  The eigenvalues of the Jacobian matrix at $(u_c,0)$ are $0$ and $-1+2u_c^2$.  Since $|u_c|\leq 1$ so $-1\leq -1+2u_c^2\leq 1$ which follows that the vector field near the critical point $(u_c,0)$ is repelling along $w$ axis if $\frac{1}{\sqrt{2}}<u_c\leq 1$ or $-1\leq u_c<-\frac{1}{\sqrt{2}}$ and attracting along $w$ axis if $-\frac{1}{\sqrt{2}}<u_c<\frac{1}{\sqrt{2}}$.   On the other hand, for $w\neq 0$ and $u>0$, we have form equation (\ref{PCS2}) that $\frac{du}{dN}>0$ and for $u<0$, we have $\frac{du}{dN}<0$.  One may note that the flow comes towards the line of critical points rapidly compare to it goes away from the critical line (see figure (\ref{PSb1})).   So the vector field near the $\left(\pm \frac{1}{\sqrt{2}},0\right)$ are also saddle in nature.   Cosmological bounce may take place near the points at infinity when the non-isolated critical points on the equator of the Poincar\'{e} sphere are saddle in nature (see section \ref{Bounce}).
		    
		\section{Cosmological Implication \label{CI}}

		\subsection{Cosmological Implication for Model $1$}
		For the critical points $C_1$ and $C_2$, as along the eigendirection of $x$, $x$ decreases to $0$ so $\dot{\phi}\rightarrow 0$.  This implies $\phi=const.$ and it indicates that the scalar field behaves as a cosmological constant.  Further, due to the presence of negative eigenvalue corresponding to $y$ coordinate, along the eigendirection of $y$, $y\rightarrow \pm 1$ which follows that $V(\phi)\rightarrow 3H^2 = Constant$.  So we have the Hubble parameter to be constant. Thus the scale factor grows exponentially.  Hence, the critical points described the quasi-de Sitter expansion in the early inflationary era.
		
		Now from the stability analysis of $C_1$ and $C_2$, we have that $z\rightarrow 0$ for $\mu>0$, $z\rightarrow \infty$ for $\mu<0$ along the eigen-direction of $z$ which implies that $\phi\rightarrow \infty$ for $\mu>0$ and $\phi\rightarrow 0$ for $\mu<0$.  Thus for $\mu<0$ the scalar field behaves as a cosmological constant and for $\mu>0$ the observation is not cosmologically significant.  
		
		For the critical points $C_3$ and $C_4$, as along the eigen-direction $x\rightarrow \infty$ so $\dot{\phi}\rightarrow \infty$ and this is not interesting from a cosmological point of view.  Further due to the presence of positive eigenvalue corresponding to $y$ coordinate along the eigendirection of $y$, $y\rightarrow \infty$ which follows that $V(\phi)\rightarrow \infty$ and this is not interesting from a cosmological point of view.  Now from the stability analysis of $C_3$ and $C_4$, we have $z\rightarrow \infty$ along the eigendirection of $z$ which implies that $\phi\rightarrow 0$ and so the scalar field behaves as the cosmological constant.

		\subsection{Cosmological Implication for Model $2$}
		
		For the critical point $P_0$ : along the eigen-direction of $x$ as $t\rightarrow \infty$, $x\rightarrow 0$ which follows that $\dot{\phi}\rightarrow 0$.  This implies $\phi=const.$ and it indicates that the scalar field behaves as a cosmological constant.  Further, along the eigen-direction of $y$ as $t\rightarrow \infty$, $y\rightarrow \infty$ which follows that $V(\phi)\rightarrow \infty$.  For the critical point $P_1$ : along the eigen-direction of $x$ as $t\rightarrow \infty$, $x\rightarrow \infty$ which follows that $\dot{\phi}\rightarrow \infty$.  Further along the eigen-direction of $y$ as $t\rightarrow \infty$, $y\rightarrow 0$ for $\nu>\sqrt{6}$ and $y\rightarrow \infty$ for $\nu<\sqrt{6}$.  It follows that $V\rightarrow 0$ for $\nu>\sqrt{6}$ and $V\rightarrow \infty$ for $\nu<\sqrt{6}$.  For the critical point $P_2$ : along the eigen-direction of $x$ as $t\rightarrow \infty$, $x\rightarrow \infty$ which follows that $\dot{\phi}\rightarrow \infty$.  Further along the eigen-direction of $y$ as $t\rightarrow \infty$, $y\rightarrow 0$ for $\nu>-\sqrt{6}$ and $y\rightarrow \infty$ for $\nu<-\sqrt{6}$.  It follows that $V\rightarrow 0$ for $\nu>-\sqrt{6}$ and $V\rightarrow \infty$ for $\nu<-\sqrt{6}$.  These results are not interesting from a cosmological point of view. 
		 
		For the critical points $P_3$ and $P_4$ : along the eigen-direction of $x$ as $t\rightarrow \infty$, $x\rightarrow \frac{\nu}{\sqrt{6}}$ if $|\nu|<\sqrt{2}$ and $x\rightarrow \infty$ if $\nu>\sqrt{2}$ or $\nu<-\sqrt{2}$.  It follows that as $t\rightarrow \infty$, $\dot{\phi}\rightarrow \nu H\implies \phi\rightarrow \nu \lim\limits_{t\rightarrow \infty} \ln a(t)\implies \phi \rightarrow \infty$ for $|\nu|<\sqrt{2}$ and also $\dot{\phi}\rightarrow \infty$ if $\nu>\sqrt{2}$ or $\nu<-\sqrt{2}$.  As the critical points $P_3$ and $P_4$ both exist for $0\leq\nu^2\leq6$, so as $t\rightarrow \infty$, $y\rightarrow \pm \sqrt{1-\frac{\nu^2}{6}}$ which follows that $V\rightarrow 3H^2\left(1-\frac{\nu^2}{6}  \right)$, that is, for $\nu=0$ it leads to an accelerated era of evolution.  Further, for the critical points $P_5$ and $P_6$ : along the eigen-directions of $x$ and $y$ as $t\rightarrow \infty$, $x\rightarrow \frac{1}{\nu}\sqrt{\frac{2}{3}}$ and $y\rightarrow \frac{2}{\sqrt{3}\nu}$ respectively while $|\nu|>\frac{2\sqrt{2}}{\sqrt{3}}$ which follows that as $t\rightarrow \infty$, $\dot{\phi}\rightarrow \frac{2}{\nu}$ and $V\rightarrow \frac{4H^2}{\nu^2}$ while $|\nu|>\frac{2\sqrt{2}}{\sqrt{3}}$.  Further for $\nu\in \left(-\sqrt{2},0\right)\cup \left(0,\sqrt{2}\right)$ as $t\rightarrow \infty$, $x\rightarrow \infty$ and $y\rightarrow \frac{2}{\sqrt{3}\nu}$ which follows that as $t\rightarrow \infty$, $\dot{\phi}\rightarrow \infty$ and $V\rightarrow \frac{4H^2}{\nu^2}$ while $\nu\in \left(-\sqrt{2},0\right)\cup \left(0,\sqrt{2}\right)$.  It leads to an accelerated phase of the universe for $\nu=\pm \frac{2}{\sqrt{3}}$.\\
	
	\begin{figure}
		\begin{subfigure}[b]{0.49\textwidth}
			\centering
			\includegraphics[width=\textwidth]{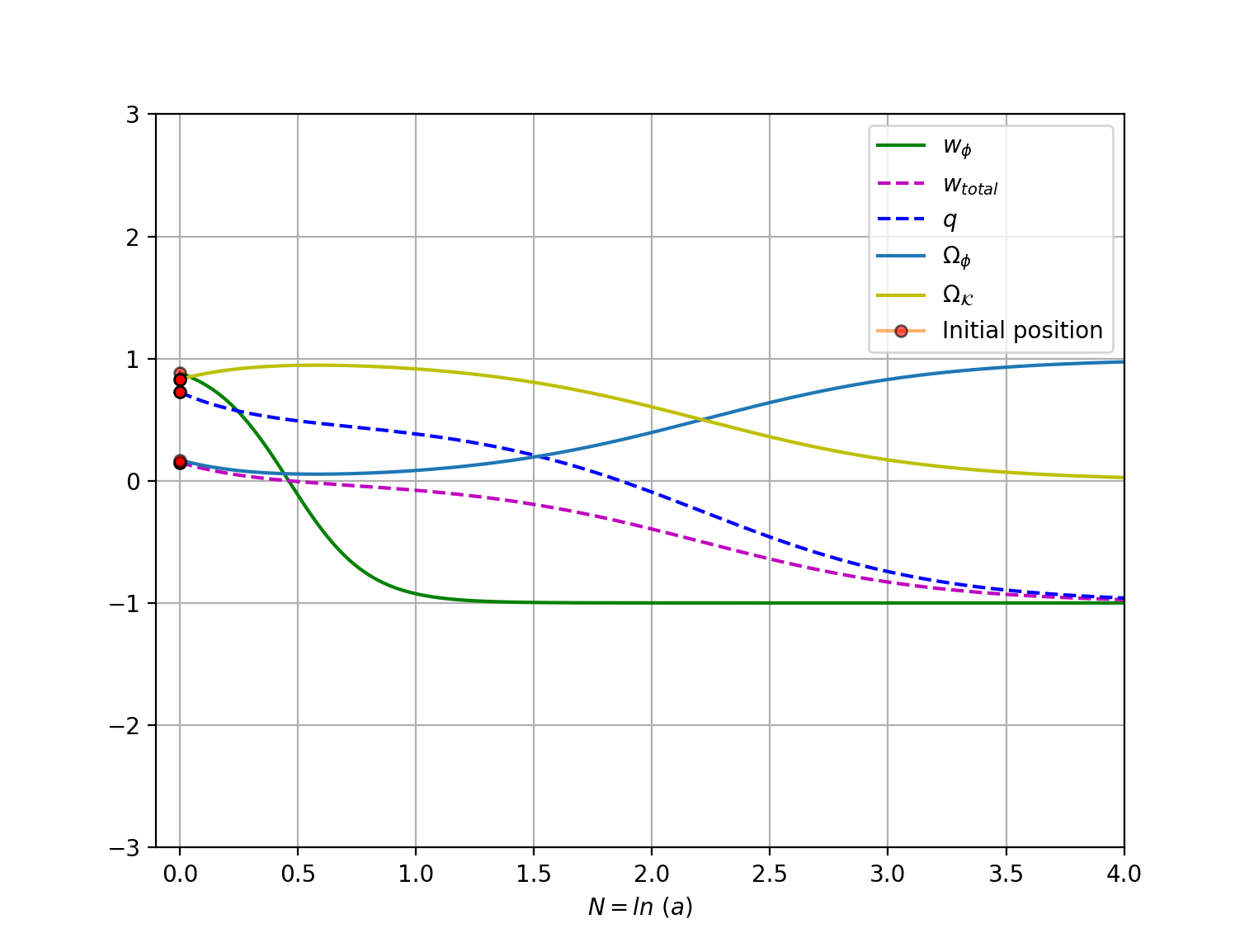}
			\caption{}
			\label{Power_Law}
		\end{subfigure}
		\hfill
		\begin{subfigure}[b]{0.48\textwidth}
			\centering
			\includegraphics[width=\textwidth]{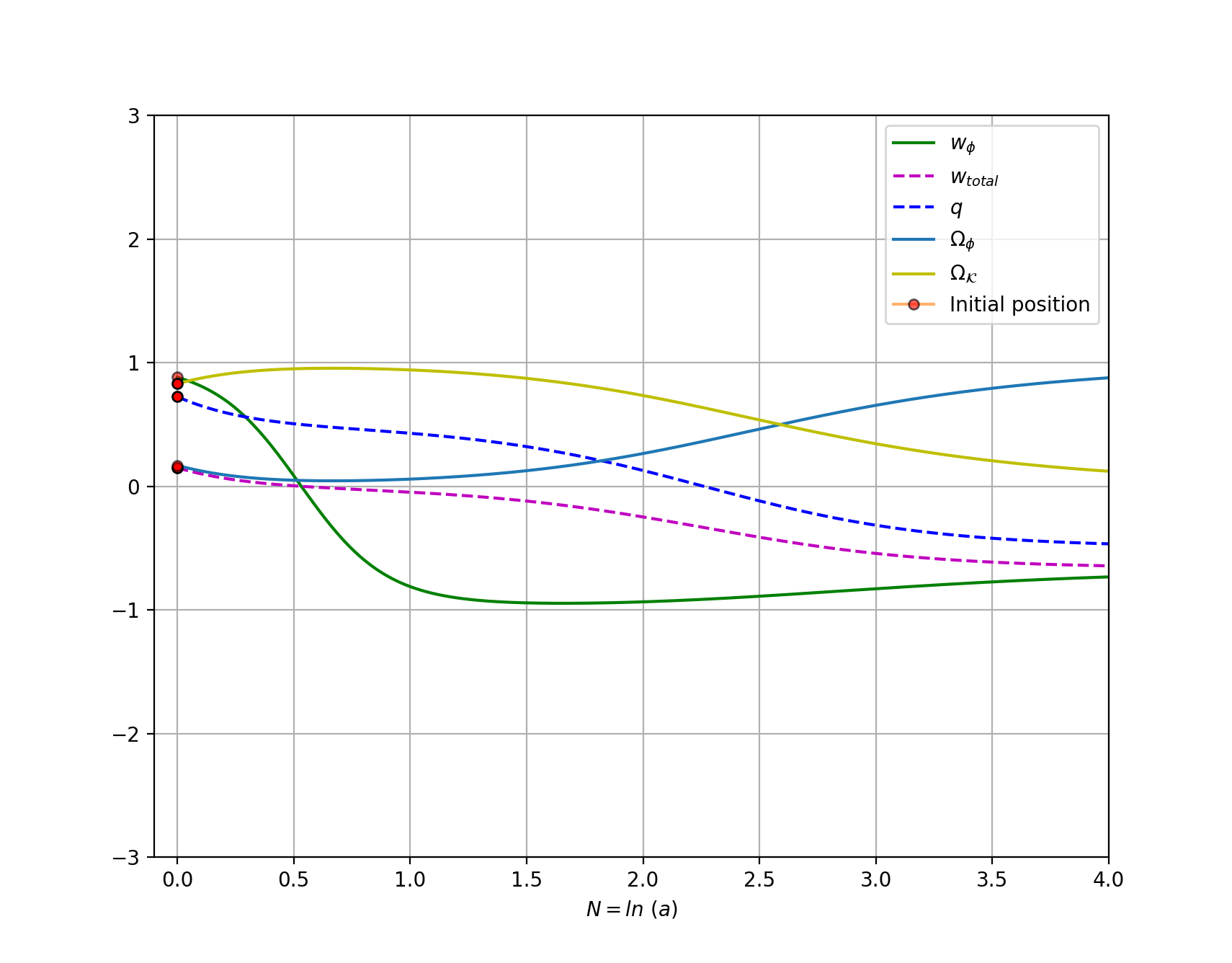}
			\caption{}
			\label{Exponential_Law}
		\end{subfigure}
	\caption{The figures show the time evolution of cosmological parameters of our cosmological model.  In panel (a), Power potential: Late time solutions of cosmological parameters.  For $\mu=1$, the kinetic energy dominated late time accelerated solutions are attracted towards the phantom boundary.  In panel (b), Exponential potential: Late time solutions of cosmological parameters.  For $\nu^2<2$, the kinetic energy dominated late time accelerated solutions are attracted towards the quintessence era which agrees with the present observed accelerated universe $(-0.56 <q<-0.49)$.} 
	\label{fig:three graphs}
\end{figure}
		
			\subsection{Cosmological Bouncing Scenarios}\label{Bounce}
			Generally speaking, the cosmological bounce takes place for  $\dot{a}\approx 0$ and $\ddot{a}>0$ or  $\ddot{a}<0$. On the other hand, one can derive that $\ddot{a}>0$ or $\ddot{a}<0$ iff $\rho_\phi+3p_\phi<0$ or $\rho_\phi+3p_\phi>0$ respectively (see Proposition (\ref{A}) in Appendix).  For $\ddot{a}>0$ the universe experiences an accelerating phase and for $\ddot{a}<0$ the universe experiences decelerating phase of evolution.  Now for $\dot{a}\approx 0$ we have $\dot{{H}}\approx \frac{\ddot{a}}{a}$.   So, in terms of Hubble parameter $H$, we can say that the bounce takes place when $H \approx 0$ and $\dot{H}>~ \text{or} ~<0$.  We also note that $\ddot{a}>~ \text{or} ~ <0$  iff $y^2>2x^2$ or $y^2<2x^2$ respectively (see Proposition (\ref{B}) in the Appendix and Table (\ref{T_Bounce})).
			
			$\bullet$ For model 1, as mentioned previously, near the critical points $C_1-C_4$ the solutions are completely dominated by scalar field and $\dot{{H}}\approx 0$ due to the condition $|\Omega_{\mathcal{K}}| \approx 0$.  So bounce does not take place if the initial conditions are given near these points.  Nevertheless, near $C_3$ and $C_4$ (unstable in nature), the spatial curvature is negligible, and $\dot{{H}}$ is negative.   As expected, we also note that $y^2<2x^2$ near these critical points and $\ddot{a}<0$ (see Table (\ref{T_Bounce})) for all values of $\mu$.  So, if the Hubble parameter takes positive sign (expanding universe) near these critical points, the expansion rate begins to decrease (as $\ddot{a}<0$) and the trajectory goes towards the critical points at infinity (Minkowski limit with $H\rightarrow 0$) in the compactified phase space (see section (\ref{Infinity})).  The trajectory soon commences to go away from those critical points (at infinity) which are saddle/saddle node in nature and $H$ becomes soon negative, i.e., the universe begins to contract. 
			
			On the other hand, near the critical point $C_5$, $\dot{H}$ is positive for positive spatial curvature.  So, if the Hubble parameter takes negative sign (contracting universe) near these critical points, the expansion rate begins to increase and the trajectory goes towards the critical points at infinity (Minkowski limit with $H\rightarrow 0$) in the compactified phase space (see section (\ref{Infinity})).  The trajectory soon commences to go away from those critical points (at infinity) which are saddle/saddle nodes in nature and contrary to the previous instance, $H$ becomes soon positive, i.e., the universe begins to expand.  The cosmological bounce takes place for $\mu>0$ when the Hubble parameter $H$ changes its sign and the trajectory asymptotically goes towards $C_1$ or $C_2$ (see table \ref{T_Bounce}).   
			
			So on the proviso that the bounce takes place, there may appear homoclinic and heteroclinic orbits due to the existence of saddle points.  So, there are several prospects to acquire various late-time solutions which depend on the initial condition; for instance, we acquire scalar field dominated time de-Sitter solution to the evolution of the universe if the trajectory asymptotically goes towards $C_1$ or $C_2$  (see figure (\ref{c_12345})) on the $XY$ plane, late time decelerating dust-dominated universe if the trajectory asymptotically goes towards $C_5$ along stable eigen direction.
			
			$\bullet$ For model 2, we have discussed earlier that near $P_0$ (saddle for all $\nu$) the kinetic term is subdominant compared to the curvature contribution.  If initially the universe is in the contracting phase then near $P_0$ the expansion rate $H$ increases for positive spatial curvature.  Depending on the initial condition if the flow goes away from $P_0$ towards infinity (as $H\rightarrow 0$) where the Hubble parameter changes its sign (when critical points at infinity are saddle/saddle nodes in nature) then the universe begins to expand.  The expansion rate decreases soon when the spatial curvature contribution becomes negligible. 
			
			On the other hand, near $P_1$ and $P_2$ (saddle-node / unstable node), the solutions are dominated by the kinetic energy of the scalar field and  $\dot{{H}}$ is negative.   As expected, it is also to be noted that $y^2<2x^2$ near these critical points and $\ddot{a}<0$ (see Table (\ref{T_Bounce})) for all values of $\nu$.   Similar to the case of $C_3$ and $C_4$ in Model 1, the cosmological bounce takes place when the Hubble parameter changes its sign at Minkowski limit with $H\rightarrow 0$, i.e., at critical points (at infinity) which are saddle/saddle-node in nature.  
			
			The non-isolated critical points $P_3$ and $P_4$ are completely dominated by kinetic energy and saddle-node in nature for $2\leq \nu^2 \leq 6$ and  $\dot{{H}}$ is negative near these points.  Unsurprisingly,  $y^2<2x^2$ near these critical points and $\ddot{a}<0$ (see Table (\ref{T_Bounce})) for $2<\nu^2 \leq 6$.  So depending on the initial condition, the cosmological bounce may take place if the flow starts near these critical points.   			
			The non-isolated critical points $P_5$ and $P_6$ are saddle in nature for $\nu^2 \leq 2$.  If $\nu \rightarrow 0$, then the critical points are dominated by spatial curvature and the sign of $\dot{{H}}$ depends on the sign of $\mathcal{K}$.  If $\nu^2 \approx  2 $, then the critical points are completely dominated by kinetic energy and $\dot{{H}}<0$.  Similar to the previous case, the initial condition plays an important role in initiating the cosmological bounce.  
			
			Similar to model 1 when it comes to the case of cosmological bounce,  there may appear homoclinic and heteroclinic orbits due to the existence of saddle points.  So, there are several prospects to acquire various late-time solutions which depend on the initial condition; for instance,  we come by late time decelerating dust-dominated universe if trajectory asymptotically approaches towards $P_0$, 
			late-time kinetic energy dominated solutions with a stiff equation of
			state if trajectory asymptotically approaches towards $P_1$ or $P_2$,
			late time kinetic energy dominated accelerated solution as trajectory asymptotically approaches towards $P_3$ or $P_4$,   late time scaling solutions where DE  behaves as quintessence boundary if trajectory asymptotically approaches towards $P_5$ or $P_6$.

		\begin{table}[!]
			\caption{\label{T_Bounce}Table shows the sign of `$\ddot{a}$' near the critical points of Model 1 and 2 in terms of dynamical variables (for details, see Proposition \ref{B} in Appendix)}. 		 
			\begin{tabular}{|c|c|c|c|}
				\hline	
				\begin{tabular}{@{}c@{}}$~~$\\ Model \\$~$\end{tabular} & Critical points & 	\begin{tabular}{@{}c@{}}$~~$\\ $y^2>2x^2 ~i.e.~ \ddot{a}>0  $\\$~$\end{tabular} &	\begin{tabular}{@{}c@{}}$~~$\\ $y^2<2x^2~i.e.~ \ddot{a}<0$\\$~$\end{tabular}\\ \hline\hline
				~&~&~&~\\
				~ & $C_1$ & Satisfied for all $\mu\neq 0$ & Not applicable\\ ~&~&~&~\\~ & $C_2$ & Satisfied for all $\mu\neq 0$ & Not applicable\\ ~&~&~&~ \\ 
				Model $1$ & $C_3$ & Not applicable & Satisfied for all $\mu$\\ ~&~&~&~\\ 	~ & $C_4$ & Not applicable & Satisfied for all $\mu$\\~&~&~&~\\ ~ & $C_5$ & Not applicable & Not applicable\\ ~ & ~ & ~ & ~ \\ \hline \hline  ~&~&~&~\\
				~ & $P_0$ & Not applicable & Not applicable\\ ~&~&~&~\\~ & $P_1$ &  Not applicable & Satisfied for all $\nu$\\ ~&~&~&~ \\ 
				~ & $P_2$ & Not applicable & Satisfied for all $\nu$\\ Model $2$&~&~&~\\ ~ & $P_3$ & Satisfied for $0\leq\nu^2<{2}$ & Satisfied for $2<\nu^2 \leq{6}$\\ ~&~&~&~ \\ ~& $P_4$ & Satisfied for  $0\leq\nu^2<{2}$ & Satisfied for $2<\nu^2 \leq{6}$ \\ ~&~&~&~\\
				~ & $P_5,P_6$ & Not applicable & Not applicable\\ ~&~&~&~\\  \hline 
			\end{tabular}
		\end{table}

		\section{Brief discussion and concluding remarks\label{conclusion}}
	
	We have studied the cosmological model with spatial curvature in the background of FLRW metric where the scalar field is minimally coupled and plays a role of dark matter content.  Two models have been discussed here.   In the first model, potential $V(\phi)$ of scalar field is taken as power-law function of scalar field $\phi$.  Whereas in the second interaction model, the potential term is taken as the exponential function of scalar field $\phi$.  Since the cosmological evolution equations are non-linear and complicated in nature, we have performed dynamical system analysis to accomplish the qualitative behavior of the cosmological model.  We have explained all possible scenarios of the bouncing universe and the corresponding late-time solutions.

We have obtained five non-hyperbolic type critical points from model 1 and from model 2 we have obtained seven critical points which are characterized (hyperbolic and non-hyperbolic) by the parameter $\nu$.  To study the nature of hyperbolic critical points we apply linear stability theory (Hartman-Grobman theorem).  On the other hand, to analyze non-hyperbolic points center manifold theory is employed to obtain the exact dynamical nature of the points on the sets.  Since at the bouncing scenarios, some variables become singular, we have performed the dynamical analysis around the critical points at infinity in section IV.  At infinity, the critical points are normally hyperbolic and saddle/saddle-node in nature (see figure (\ref{PS1})).  Model 1 can not alleviate the cosmic coincidence problem whereas model 2 can take the edge off the cosmic coincidence problem depending on the parameter $\nu$.

In Model $1$, there are five non-hyperbolic equilibrium points.  The first two equilibrium points $C_1$ and $C_2$ describe the flat model of the universe with a cosmological constant and they describe the phantom barrier of accelerated expansion.  The point $C_1$ represents an expanding model while $C_2$ corresponds to a contracting model of the universe.  The other three equilibrium points $C_3$, $C_4$, $C_5$ represent a decelerating phase of expansion.  From a cosmological point of view only equilibrium point $C_5$ is interesting, it describes a dust era of evolution.  Further, the critical points $C_1$ and $C_2$ are purely DE dominant $(\Omega_\phi=1)$ and they are analogous to the observationally favored $\Lambda$CDM model.  The critical points $C_3$ and $C_4$ are not of much interest.  Here the result matter behaves as stiff fluid which is effective at the very early era when quantum effects are important.  The last critical point $C_5$ has no effect of DE $(\Omega_\phi=0,~ \Omega_\mathcal{K}=1)$ and the resulting fluid behaves as dust.  

In Model $2$, there are seven equilibrium points all of which are non-hyperbolic in nature.  The dust era of decelerated expansion is characterized by the equilibrium point $P_0$.  The equilibrium points $P_1$ and $P_2$ are not interesting from a cosmological point of view.    The other four equilibrium points $P_3$, $P_4$, $P_5$, and $P_6$ represent scaling cosmological solutions.  The equilibrium points $P_3$ and $P_4$ describe a flat FLRW model with an accelerating phase of expansion for $\nu^2<2$ and the scalar field describes a dark energy model.  The equilibrium points $P_5$ and $P_6$ also describe the decelerating era of expansion for $\nu^2>2$ and here curvature contribution of the matter is non-zero and the scalar field behaves as perfect fluid at the quintessence barrier.  Further, if $\nu>2$ then curvature matter contribution will dominate over the scalar field while if $\sqrt{2}<\nu<2$ then the scalar field is the dominant matter component.  Moreover, from cosmological view point the critical point $P_0$ describes a dust era of evolution having no effect of DE $(\Omega_\phi=0)$.  Due to unacceptable cosmological parameters the critical points $P_1$ and $P_2$ may be discarded.  The critical points $P_3$ and $P_4$ are completely dominated by DE $(\Omega_\phi=1)$.  For $0\leq \nu^2\leq 2$ the present model describes the late time accelerated phase (as predicted by observational data) and $\nu=0$ describes the observationally supported $\Lambda$CDM model (as expected $V(\phi)$ becomes $V_0$, a constant), while for $2<\nu^2\leq 3$ the model goes back to the decelerated era of evolution with $\nu^2=3$ represents the dust epoch.   Lastly the critical points $P_5$ and $P_6$ have similar behavior as $P_3$ and $P_4$ and they describe the $\Lambda$CDM model with the choice $\nu^2=2/3$.  Thus it is possible to have an analogy of the present models with the well-known $\Lambda$CDM cosmology.  Moreover, if $\nu^2<2/3$ then the critical points $P_5$ and $P_6$ may go beyond $\Lambda$CDM and they may be related to other tensions in cosmology. Though both the potential models are cosmologically viable yet model $2$ has an edge over model $1$.  In model $2$, there are scaling solutions (corresponding to critical points $P_3-P_6$) indicating the possibility of decelerating era of expansion to the present accelerated expansion phase.  However, such type of expansion is not possible for power law of potential.

Thus, in summary, the present cosmological model based on the spatial curvature with power law and exponential potential may describe different evolutionary phases of the universe for which bounce takes place.  To study the bouncing universe we need to analyze the behavior of the vector field at the Minkowski limit with $H\rightarrow 0$, i.e., at critical points (at infinity) which are saddle/saddle-node in nature.  Depending on the initial condition, either the transition comes to pass from expansion to contraction and vice versa, from contraction to expansion of the universe. Although, so far,  there is neither any observational evidence that substantiates   the bouncing nature of the Universe nor any evidence of contracting phase of evolution.

  The present model describes a change in the universe's expansion rate (i.e., from expansion to contraction) so it may be possible to have an explanation or it may alleviate the standard $H_0$-tension.  The uncoupled scalar field models do not really increase the $H_0$ value \cite{Planck:2018vyg,Yang:2018xah}. Though a coupled scalar field model can reduce the $H_0$ tension, a proper choice of the coupling function between the scalar field and dark matter \cite{Gomez-Valent:2022bku}, it is possible to increase the $H_0$ value compared to the $\Lambda$CDM based Plank's estimation \cite{Planck:2018vyg}.  As a result, the obtained value of $H_0$ could be closer to the SH0ES value \cite{Riess:2021jrx}.  One may note that the inclusion of curvature in the cosmological models does not offer any solution to the Hubble constant tension, rather the tension on $H_0$ increases.  Therefore, even if the consideration of the curvature of the Universe gives a complete picture of the underlying cosmological model, however, the increasing tension in $H_0$ needs special attention with the growing sensitivity in the upcoming astronomical probes.  In addition, two other additional tensions related to $H_0$-tension in cosmology, namely the $\sigma_8$ tension and the galaxy rotation curves problem may have some issues which we shall study in a future work.

		\begin{acknowledgements}
			The authors thank the anonymous referees whose comments and suggestions improved the quality of the paper.  Soumya Chakraborty thanks CSIR, Govt. of India for providing Senior Research Fellowship (CSIR Award No: 09/096(1009)/2020-EMR-I) for the Ph.D. work.\\
		\end{acknowledgements}
		
		Authors' comment: This article describes entirely theoretical research. So data sharing is not applicable to this article as no datasets were generated or analyzed during the current study.\\
		
		Conflict of Interest: The authors declare that they have no conflict of interest.\\\\

		\section{Appendix}	

\begin{prop}
	If $\rho_\phi$ is the energy density and $p_\phi$ is the pressure of the scalar field (given by equations (\ref{eq5}) and (\ref{eq6})) then the second order derivative of the scale factor $a(t)$ with respect to the cosmic time $t$, that is, $\ddot{a}>0$ or $\ddot{a}<0$ iff $\rho_\phi+3p_\phi<0$ or $\rho_\phi+3p_\phi>0$ respectively. \label{A}
\end{prop}
\textsc{Proof:} The expression of Hubble parameter $H(t)$ in terms of scale factor $a(t)$ is defined by
$$
H=\frac{\dot{a}}{a}.
$$
Taking differentiation on both sides of the above with respect to $t$ yields
$$
\dot{H}=\frac{\ddot{a}}{a}-\frac{\dot{a}^2}{a^2}.
$$
Substituting $H=\frac{\dot{a}}{a}$ in left hand side of Eq.(\ref{eq7}) yields
\begin{align}
	3\frac{\dot{a}^2}{a^2}&=\frac{1}{2}\dot{\phi}^2+V(\phi)-3\frac{\mathcal{K}}{a^2},\nonumber\\    \frac{\dot{a}^2}{a^2}&=\frac{1}{3}\left(\frac{1}{2}\dot{\phi}^2+V(\phi)-3\frac{\mathcal{K}}{a^2}\right).\label{A1}
\end{align}
Plugging the values of $\dot{H}$ in left hand side of Eq.(\ref{eq8}) and using (\ref{A1}) yields
\begin{align}
	2\left(\frac{\ddot{a}}{a}-\frac{\dot{a}^2}{a^2}\right)&=-\dot{\phi}^2+2\frac{\mathcal{K}}{a^2},\nonumber \\   2\frac{\ddot{a}}{a}&=2\frac{\dot{a}^2}{a^2}-\dot{\phi}^2+2\frac{\mathcal{K}}{a^2},\nonumber \\   2\frac{\ddot{a}}{a}&=\frac{2}{3}\left(\frac{1}{2}\dot{\phi}^2+V(\phi)-3\frac{\mathcal{K}}{a^2}\right)-\dot{\phi}^2+2\frac{\mathcal{K}}{a^2},\nonumber \\   2\frac{\ddot{a}}{a}&=\frac{1}{3}\dot{\phi}^2+\frac{2}{3}V(\phi)-\dot{\phi}^2, \nonumber\\    \frac{\ddot{a}}{a}&=-\frac{1}{3}(\dot{\phi}^2-V(\phi)).\label{A2}
\end{align}
From equations (\ref{eq5}) and (\ref{eq6}), we have
$$
\begin{aligned}
	\rho_\phi&=\frac{1}{2}\dot{\phi}^2+V(\phi),\\ p_\phi&=\frac{1}{2}\dot{\phi}^2-V(\phi).
\end{aligned}
$$
Note that
$$
\rho_\phi+3p_\phi=\frac{1}{2}\dot{\phi}^2+V(\phi)+\frac{3}{2}\dot{\phi}^2-3V(\phi)=2(\dot{\phi}^2-V(\phi))
$$
which implies
$$
\dot{\phi}^2-V(\phi)=\frac{1}{2}(\rho_\phi+3p_\phi).
$$
Using the above result on the right-hand side of (\ref{A2}) yields
\begin{equation}
	\frac{\ddot{a}}{a}=-\frac{1}{6}(\rho_\phi+3p_\phi). \label{RE}
\end{equation}
Now one may note from the equation (\ref{RE}) that $\ddot{a}>0$ iff 
$$
\begin{aligned}
	-\frac{1}{6}(\rho_\phi+3p_\phi)&>0, \\   \rho_\phi+3p_\phi&<0.
\end{aligned}
$$

and $\ddot{a}<0$, iff
$$
\begin{aligned}
	\rho_\phi+3p_\phi&>0.
\end{aligned}
$$

This completes the proof.\\

\begin{prop}
	{The second order derivative of the scale factor $a(t)$ with respect to the cosmic time $t$, that is, $\ddot{a}>0$ or $\ddot{a}<0$ iff $y^2>2x^2$ or $y^2<2x^2$ respectively where $x$ and $y$ are the dynamical variables given by the equations (\ref{eqn11}) and (\ref{eqn12}).\label{B}}
\end{prop}
\textsc{Proof:} In the previous proposition, we have shown that at the point of bouncing condition
$$
\rho_\phi +3p_\phi<0~\text{or}~\rho_\phi+3p_\phi>0.
$$
From the equations (\ref{eq5}) and (\ref{eq6}), we have
$$
\begin{aligned}
	\rho_\phi&=\frac{1}{2}\dot{\phi}^2+V(\phi),\\ p_\phi&=\frac{1}{2}\dot{\phi}^2-V(\phi).
\end{aligned}
$$
Plugging the expressions of $\rho_\phi$ and $p_\phi$ into the first bouncing condition (i.e., $\rho_\phi+3p_\phi<0$) yields
\begin{align}
	\frac{1}{2}\dot{\phi}^2+V(\phi)+3\left(\frac{1}{2}\dot{\phi}^2-V(\phi)\right)&<0 ,\nonumber \\   2\dot{\phi}^2-2V(\phi)&<0, \nonumber\\   \dot{\phi}^2&<V(\phi).\label{B1}
\end{align}
In terms of dynamical variables, from equations (\ref{eqn11}) and (\ref{eqn12}), we have
$$\begin{aligned}
	\dot{\phi}&=\sqrt{6}Hx,\\V(\phi)&=3H^2y^2.
\end{aligned}
$$
Substituting these into (\ref{B1}) yields
$$
\begin{aligned}
	(\sqrt{6}Hx)^2&< 3 H^2 y^2, \\   6H^2 x^2 &< 3H^2 y^2, \\   y^2&>2x^2.
\end{aligned}
$$ 
Similarly, by substituting the values of $p_\phi$ and $\rho_\phi$ from the equations (\ref{eq5}) and (\ref{eq6}) into the second bouncing condition (i.e., $\rho_\phi+3p_\phi>0$), we can show that $y^2<2x^2$.  This completes the proof.\\

\begin{prop}
	{Equation (\ref{eq17}) can be obtained by using equations (\ref{eq14}), (\ref{eq15}) and the constraint equation (\ref{eq23}).\label{C}}
\end{prop}

\textsc{Proof:}  By taking differentiation both sides of (\ref{eq23}) with respect to $N$ yields
$$
\begin{aligned}
	&2x\frac{dx}{dN}+2y\frac{dy}{dN}-6\mathcal{K}u\frac{du}{dN}=0, \\  &\frac{du}{dN}=\frac{1}{3\mathcal{K}u}\left\{x\frac{dx}{dN}+y\frac{dy}{dN}\right\}.
\end{aligned}
$$
Now by substituting the equations (\ref{eq14}) and (\ref{eq15}) in the right hand side of above yields
$$
\begin{aligned}
	\frac{du}{dN}&=\frac{1}{3\mathcal{K}u}\left\{-3x^2-\sqrt{\frac{3}{2}}\left(\frac{V'}{V}\right)xy^2+3x^4-3\mathcal{K}u^2x^2+3y^2x^2-3\mathcal{K}u^2y^2+\sqrt{\frac{3}{2}}\left(\frac{V'}{V}\right)xy^2\right\}\\&=\frac{1}{3\mathcal{K}u}\{-3\mathcal{K}u^2(x^2+y^2)+3x^2(y^2+x^2-1)\}.
\end{aligned}
$$
From the equation (\ref{eq23}), we have
$$
x^2+y^2-1=3\mathcal{K}u^2.
$$
Plugging this into the second term of the right-hand side of the last expression yields
$$
\begin{aligned}
	\frac{du}{dN}&=\frac{1}{3\mathcal{K}u} \{9\mathcal{K}u^2x^2-3\mathcal{K}u^2(x^2+y^2)\}\\&=\frac{1}{3\mathcal{K}u}\{3\mathcal{K}u^2(3x^2-x^2-y^2)\}\\&=u(2x^2-y^2).
\end{aligned}
$$
This completes the proof.\\

\begin{prop}
	{The expressions of the center manifold corresponding to the critical point $C_1$ are given by $u=0$ and $v=-\frac{\mu^2}{72}w^2+\mathcal{O}(w^4)$, and the flow on the center manifold is determined by $\frac{dw}{dN}=-\frac{\mu}{6}w^3+\mathcal{O}(w^4)$.\label{CPC1}}
\end{prop}

\textsc{Proof:} The Jacobian matrix at the critical point $C_1$ can be put as
\begin{equation}	
	J(C_1)=\renewcommand{\arraystretch}{1.5}\begin{bmatrix}
		-3&0&\frac{\mu}{2}\\0&-2&0\\0&0&0
	\end{bmatrix}.\label{eq36}	
\end{equation}
The eigenvalues of the above matrix are $-3$, $-2$ and $0$ with $[1, 0, 0]^T$, $[0, 1, 0]^T$ and $\left[\frac{\mu}{6}, 0, 1\right]^T$ are the corresponding eigenvectors respectively.  To apply center manifold theory, first, we transform the coordinates into a new system $x=X,~ y=Y+1,~ z=Z$, such that the critical point $C_1$ moves to the origin. By using the eigenvectors of the Jacobian matrix $J(C_1)$, we introduce another set of new coordinates $(u,~v,~w)$ in terms of $(X,~Y,~Z)$ as
\begin{equation}\renewcommand{\arraystretch}{1.5}	
	\begin{bmatrix}
		u\\
		v\\
		w
	\end{bmatrix}\renewcommand{\arraystretch}{1.5}
	=\begin{bmatrix}
		1 & 0 & -\frac{\mu}{6} \\	
		0 &  1 & 0 \\
		0 & 0 & 1
	\end{bmatrix}\renewcommand{\arraystretch}{1.5}
	\begin{bmatrix}
		X\\
		Y\\
		Z
	\end{bmatrix}\label{eq24}
\end{equation}		
and in these new coordinate system, the equations $(\ref{eqn29}-\ref{eqn31})$ are transformed into	
\begin{equation}	\renewcommand{\arraystretch}{1.5}
	\begin{bmatrix}
	\frac{du}{dN}\\
	\frac{dv}{dN}\\
	\frac{dw}{dN}
	\end{bmatrix}
	=\begin{bmatrix}
		-3 & ~~0 & 0 \\	
		~~0 & -2 & 0 \\
		~~0 & ~~0 & 0
	\end{bmatrix}
	\begin{bmatrix}
		u\\
		v\\
		w
	\end{bmatrix}		
	+	
	\begin{bmatrix}
		non\\
		linear\\
		terms
	\end{bmatrix}.	
\end{equation}	
By center manifold theory there exists a continuously differentiable function 	$h:\mathbb{R}$$\rightarrow$$\mathbb{R}^2$ such that 
\begin{align}\renewcommand{\arraystretch}{1.5}
	h(w)=\begin{bmatrix}
		u \\
		v \\
	\end{bmatrix}
	=\begin{bmatrix}
		a_1w^2+a_2w^3 +\mathcal{O}(w^4)\\
		b_1w^2+b_2w^3 +\mathcal{O}(w^4) 
	\end{bmatrix}.
\end{align}
Differentiating both sides with respect to $N$, we get 
\begin{eqnarray}
\frac{du}{dN}&=&(2a_1w+3a_2w^2)\frac{dw}{dN}+\mathcal{O}(w^3)\label{eq25}\\
\frac{dv}{dN}&=&(2b_1w+3b_2w^2)\frac{dw}{dN}+\mathcal{O}(w^3)\label{eq26}
\end{eqnarray}
where $a_i$, $b_i$ $\in\mathbb{R}$.  We are only concerned about the non-zero coefficients of the lowest power terms in CMT as we analyze the stability in an arbitrarily small neighborhood of the origin.  Comparing coefficients corresponding to power of $w$ both sides of (\ref{eq25}) and (\ref{eq26}), we get	
$a_i=0$ and $b_1=\frac{\mu^2}{72}$, $b_2=0$.  So, the center manifold can be written as 
\begin{eqnarray}
	u&=&0,\label{eqn27}\\
	v&=&-\frac{\mu^2}{72}w^2+\mathcal{O}(w^4)\label{eqn28}
\end{eqnarray} 
and the flow on the center manifold is determined by
\begin{eqnarray}
	\frac{dw}{dN}&=&-\frac{\mu}{6}w^3+\mathcal{O}(w^4) .\label{eq29}
\end{eqnarray}

		\bibliographystyle{unsrt}
		\bibliography{references}\bigbreak

	\end{document}